\documentclass[onefignum,onetabnum]{siamart171218}
\usepackage{lipsum}
\usepackage{cite}
\usepackage{amsfonts}
\usepackage{graphicx}
\usepackage{epstopdf}
\usepackage{amsmath}
\usepackage{algorithmic}
\usepackage{caption, subcaption}
\usepackage{pifont}
\usepackage{booktabs}
\usepackage{multirow}
\usepackage{xcolor}
\ifpdf
  \DeclareGraphicsExtensions{.eps,.pdf,.png,.jpg}
\else
  \DeclareGraphicsExtensions{.eps}
\fi
\usepackage{epstopdf}
\AppendGraphicsExtensions{.tif}

\newsiamremark{remark}{Remark}
\newsiamremark{hypothesis}{Hypothesis}
\crefname{hypothesis}{Hypothesis}{Hypotheses}
\newsiamthm{claim}{Claim}
\headers{Persistent Homology of Geospatial Data: A Case Study with Voting}{Michelle Feng, Mason A. Porter}

\title{Persistent Homology of Geospatial Data: A Case Study with Voting%\thanks{Submitted to the editors DATE.
}

\author{Michelle Feng\thanks{Department of Mathematics, University of California Los Angeles
  (\email{mhfeng@math.ucla.edu}, \url{https://math.ucla.edu/\string~mhfeng/}).}
\and Mason A. Porter\thanks{Department of Mathematics, University of California Los Angeles
  (\email{mason@math.ucla.edu}, \url{https://www.math.ucla.edu/\string~mason/}).}}

\usepackage{amsopn}

\newcommand{\xmark}{\ding{55}}

\ifpdf
\hypersetup{
  pdftitle={},
  pdfauthor={M. Feng, M.~A.~Porter}
}
\fi

\begin{document}
\maketitle

% REQUIRED
\begin{abstract}
A crucial step in the analysis of persistent homology is the transformation of data into an appropriate topological object (in our case, a simplicial complex). Modern packages for persistent homology often construct Vietoris--Rips or other distance-based simplicial complexes on point clouds because they are relatively easy to compute. We investigate alternative methods of constructing these complexes and the effects of making associated choices during simplicial-complex construction on the output of persistent-homology algorithms. We present two new methods for constructing simplicial complexes from two-dimensional geospatial data (such as maps). We apply these methods to a California precinct-level voting data set, demonstrating that our new constructions can capture geometric characteristics that are missed by distance-based constructions. Our new constructions can thus yield more interpretable persistence modules and barcodes for geospatial data. In particular, they are able to distinguish short-persistence features that occur only for a narrow range of distance scales (e.g., voting behaviors in densely populated cities) from short-persistence noise by incorporating information about other spatial relationships between precincts.
\end{abstract}

%%%%%%

%%%%%

\section{Introduction}
\label{sec:intro}

Historically, the study of algebraic topology has been concerned with classifying topological spaces based on global properties using algebraic invariants \cite{hatcher2002algebraic}. More recently, however, ideas from algebraic topology have also been applied to data sets as a way of examining the ``shape'' of data \cite{edelsbrunner2010,ghrist2008,otter2017}. One way to classify topological spaces is to distinguish them based on their number and types of holes. For example, a circle is distinct from a disc; we distinguish them based on the hole in the center of a circle. For two-dimensional (2D) geospatial\footnote{Following the conventions of the demography community, we use the term ``geospatial data'' to refer to information about entities on or near Earth's surface that one can locate using some coordinate system (in our case, using latitude and longitude). In this paper, we use the term ``geospatial'' interchangeably with ``geographic'', in contrast to more general spatial data, which need not be based on geographic location.} data, we can interpret holes as concrete geographical features like lakes or deserts.  Some previous geospatial and spatial applications of topological data analysis (TDA) include studies of the geography of country development \cite{banman2018}, the spread of social \cite{taylor2015} and biological \cite{10.1371/journal.pone.0192120} contagions, communication patterns in cities \cite{Bajardi2015}, voting in the ``Brexit referendum'' \cite{stolz-brexit}, continuum disk percolation in 2D \cite{speidel2018}, granular materials \cite{lia2018}, flow networks in biological transport \cite{flow2019}, and migration networks \cite{Ignacio2019}.

To identify holes in a data set, we need to assign a topological structure to the data and compute its homology groups. The homology of a topological space $X$ is a set of topological invariants that are represented by homology groups $\{H_k(X)\}_{k \in \mathbb{N}}$, where $H_k(X)$ describes $k$-dimensional holes in $X$. If $X$ is a network, the dimension of $H_0(X)$ records the number of connected components and the dimension of $H_1(X)$ records cycles. Because of the 2D nature of our focal data, we mostly restrict ourselves to computing these homology groups. Homology is particularly useful as a topological invariant because of the existence of efficient combinatorial algorithms for computing the homology groups of simplicial and cellular complexes \cite{otter2017}, as other topological invariants (such as homotopy) are less computationally tractable.

Within the subject of TDA, persistent homology (PH) is the most common method for computing holes in data. Point clouds have an inherent $0$-dimensional (0D) structure, and they thus have few interesting topological properties when considered simply as a finite collection of points. However, by turning a point cloud into a higher-dimensional simplicial complex, we can gain more information about its ``shape''. In PH, we take a point cloud and turn it into a series of simplicial complexes at different scales; we then compute the homology of each of these complexes, tracking homological features across scales. Intuitively, consider looking at a point cloud and filling in the areas between points that are close to each other to get a manifold. Changing our notion of what it means to be ``close to each other'' results in a collection of different manifolds, each of which approximates our original point cloud. Most simply, we can take ``close'' to mean within some Euclidean distance, and we can then progressively increase this distance. Imagine squinting at a point cloud until it blurs and takes on some shape; the harder you squint, the more the edges of the shape blur and expand. This approach is particularly useful because of its ability to encode geometric information due to the inclusion of a scaling parameter. Although topological invariants are useful because of their mathematically rigorous meaning, our intuition about what it means for a data set to have a certain shape includes many concepts that cannot be captured up to homeomorphism. Consider the classical example of a coffee cup and a donut: their homology groups are indistinguishable, yet we may still be interested in identifying differences between them.

Persistent homology has been used in a large variety of problems in numerous disciplines \cite{otter2017}. Applications of PH have included studies in protein compressibility \cite{Gameiro2015, doi:10.1002/cnm.2655, kovacev2016, doi:10.1093/bib/bbs077}, DNA structure \cite{Emmett:2016:MTC:2954721.2954838}, computer vision \cite{Carlsson2008}, a wealth of different topics in neuroscience \cite{bendich2016,curto2016, 10.7554/eLife.03476, dotko2016, Giusti2016, Kanari2016QuantifyingTI, 10.3389/fnsys.2016.00085, YOO20161}, and more. One aspect of PH that makes it very appealing is its robustness to noise: because one examines data at multiple scales simultaneously, conventional wisdom suggests that features that persist over a variety of scales should be the result of a signal (rather than of noise). However, in some data sets (as in the case of geographical data sets), several distance scales are represented within a single point cloud, making it difficult to find persistent features. More generally, for both spatial and non-spatial data, some non-persistent features can convey important signals, as illustrated in \cite{doi:10.1063/1.4978997} in an application to neuroscience. In these situations, it can be difficult to distinguish between (1) features that are real but appear only at specific scales and (2) noise.

Data sets that exhibit interesting features at multiple scales are a particularly poor fit for constructions that use distances to turn point clouds into complexes. However, distance-based constructions---especially the Vietoris--Rips (VR) construction---are the most common choice for constructing simplicial complexes from point clouds because of their relatively fast computation times \cite{Zomorodian_2010,otter2017}. Much of the recent literature on methods for constructing simplicial complexes has focused either on finding faster ways to build VR complexes or on building approximations to a VR complex using less data and thereby reducing computation time. Computing a VR complex or other distance-based complexes have been very effective for many applications \cite{otter2017}, but they can sometimes lead to considerable difficulty in interpreting results, especially in applications where scaling is a major factor. To mitigate the effect of scaling, we propose the construction of simplicial complexes based on the network or contiguity properties of an underlying data set when available, as this allows an interpretation of persistence that does not rely on distance scaling and which is thus easier to interpret for geographical data.

Our paper proceeds as follows. In Section~\ref{sec:bg}, we discuss background information about our data set and the methods of PH. In Section~\ref{sec:methods}, we discuss several methods for the construction of simplicial complexes, including traditional distance-based constructions (VR and alpha complexes) and two new constructions that are based on the contiguity of geographical data. We also discuss the differences in geometry between these methods and our intuition about how those differences affect our analysis. In Section~\ref{sec:results}, we give some computational results that support our intuition and provide guidelines for when each method of construction is appropriate. In Section~\ref{sec:conclusion}, we discuss future directions for the computation of PH on 2D data. In appendices, we give further background about simplicial homology and additional details about some of our computations and results.

%%%%

\section{Background}
\label{sec:bg}

%%%%

\subsection{Voting Data}
\label{ss:bgdata}

\begin{figure}[!htbp]
\centering
\includegraphics[height=.20\textwidth]{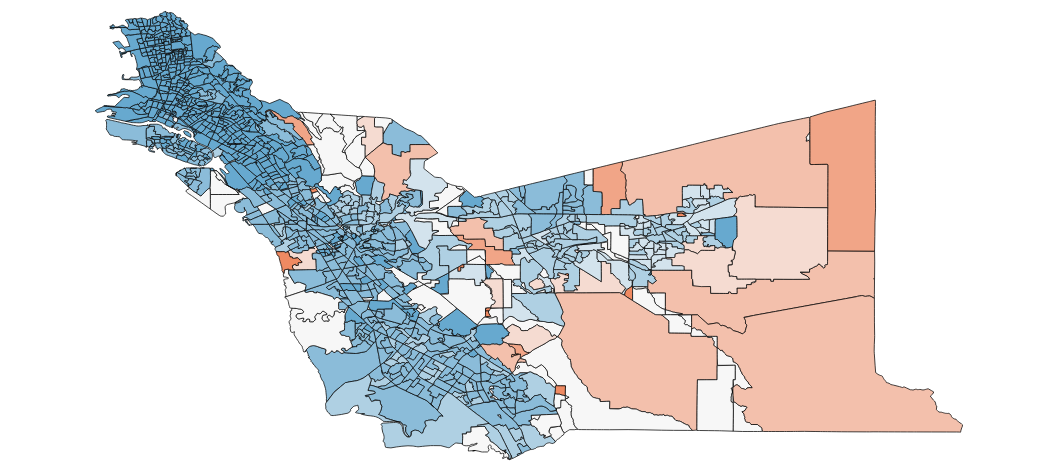}
\includegraphics[height=.20\textwidth]{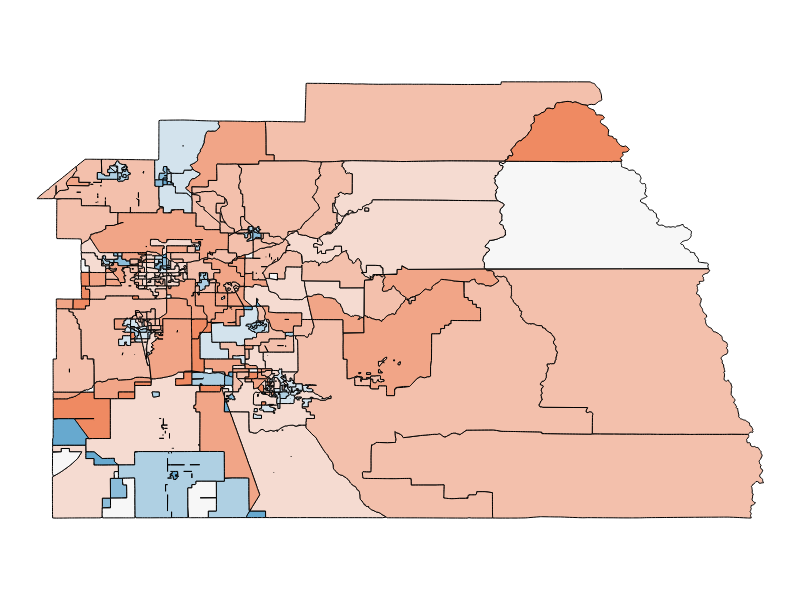}
\caption{The counties of (left) Alameda and (right) Tulare. Red precincts voted predominantly for Donald Trump, and blue ones voted predominantly for Hillary Clinton. Darker shading in a precinct indicates a stronger majority for the winning candidate, so Trump won dark-red precincts by a large margin and Clinton won dark-blue precincts by a large margin. We use the color white for precincts with a strictly equal number of votes for the two candidates.
}
\label{fig:2_dataset}
\end{figure}

Throughout our paper, we use data from the \emph{LA Times} California 2016 Election Precinct Maps \cite{schleuss2016}. Compiled by the \emph{Los Angeles Times} Data Visualization Team after the November 2016 elections, this data set has precinct-level results for all of California for every statewide race. Specifically, this encompasses results for the presidential race, California's senatorial race, and 17 statewide propositions. The data covers all of California's 24626 precincts (which are organized into $58$ counties); for each one, it includes the number of votes for each choice in each race, along with an associated {\sc shapefile} and other metadata. We generate precinct maps for each county, and we classify precincts in the presidential race based on the margin of victory for each candidate. We show two examples in Figure~\ref{fig:2_dataset}.

We are interested especially in examining the phenomenon in which a precinct votes differently from the areas that surround it (e.g., ``an island of red voters in a sea of blue'', or vice versa). Understanding this phenomenon gives one way of quantifying the voting behaviors of counties at large: some counties have rather uniform voting patterns; whereas others may contain clusters of communities that vote unlike their neighbors, potentially signaling the presence of urban areas, demographically distinct neighborhoods, or gerrymandering. 

This application is particularly well-suited to PH, because we can interpret counties with distinct voting preferences as holes. Additionally, computing the homologies of these counties allows us to classify them based on their topological features. We can consider a county as a point cloud, where each precinct is a point with some additional data assigned to it (specifically, voting preferences and the geographical space that it occupies). For the remainder of this paper, we consider only votes for the candidates Hilary Clinton and Donald Trump in the presidential election. We use ``red'' to indicate a voting preference for Trump, and we use ``blue'' to indicate a preference for Clinton, with darker colors signifying a stronger voting preference in a particular precinct.

%%%%%%

\subsection{Persistent Homology}
\label{ss:bgph}

We now give a more rigorous discussion of some of the intuitive descriptions of PH from Section~\ref{sec:intro}. Suppose that we have experimental data $X$, from which we have constructed a sequence $X_1 \subset X_2\subset \cdots\subset X_l $ of simplicial complexes of dimension $d$. In Section~\ref{sec:methods}, we will discuss several methods to construct such a sequence. We require that the sequence $\{X_i\}$ is increasing, such that it forms a filtered simplicial complex; and we call each $X_i$ a subcomplex. The filtered simplicial complex, along with inclusion maps between subcomplexes and chain and boundary maps of each subcomplex, is called a ``persistence complex''. We examine the homology of each subcomplex, noting that the inclusion map $X_i \hookrightarrow X_j$ induces a map $f_{i,j}: H_m(X_i) \to H_m(X_j)$, and that, by functoriality,
\begin{equation}
	f_{k,j} \circ f_{i,k} = f_{i,j}  \,.
\end{equation}

\begin{definition}
Let $X=X_0 \subset X_1\subset \cdots \subset X_l$ be a filtered simplicial complex. The {\bf $m$th persistent homology} of $X$ is the pair
\begin{equation*}
	\left(\left\{H_m(K_i)\right\}_{1\leq i \leq l}, \left\{f_{i,j}\right\}_{1\leq i \leq j \leq l}\right) \,,
\end{equation*}
where $f_{i,j}: H_m(X_i) \to H_m(X_j)$, for all $i \leq j$ and $m$ smaller than the dimension\footnote{Note that $m$ need not necessarily be less than the dimension of $X$, as shown in \cite{adamaszek_adams_2017}, but this is a convenient simplification for many applications (including ours).} of $X$, are the maps that are induced by the action of the homology functor on the inclusion maps $X_i \hookrightarrow X_j$. We refer to the collection of all $m$th persistent homologies as the {\bf persistent homology (PH)} of $X$.
\end{definition}

Most notably, the PH of a filtered simplicial complex encodes information about the maps between each subcomplex, thereby giving more information than the homologies of the individual subcomplexes. Each homology group with field coefficients $H_m(X_i)$ is a vector space whose generators correspond to holes in $X_i$, and the maps $f_{i,j}$ allow us to track these generators from $H_m(X_i)$ to $H_m(X_j)$. By picking a convenient basis for $H_m(X_i)$, which we are able to do by the Fundamental Theorem of Persistent Homology \cite{Zomorodian2005}, we can construct a well-defined and unique collection of disjoint half-open intervals, where each generator $x\in H_p(K_i)$ corresponds to an interval $[b_x,d_x)$, with $X_{b_x}$ denoting the subcomplex in which the generator (and its associated hole) first appears and $X_{d_x}$ denoting the subcomplex in which the generator dies. More precisely, we say that $x\neq 0 \in H_p(X_{b_x})$ is born in $X_{b_x}$ if it is not in the image of $f_{b_x -1, b_x}$; it dies in $H_p(X_{d_x})$ if $d_x > b_x$ is the smallest index for which $f_{b_x,d_x}(x) = 0$. If $f_{b_x,j}(x) \neq 0$ for all $b_x < j \leq l$, then $x$ lives forever and we associate the interval $[b_x,\infty)$ to it. For a more in-depth discussion of PH and other homological concepts, see~\cref{app:simplicial}; for further material, see \cite{hatcher2002algebraic, otter2017, ghrist2008, Zomorodian2005}.

The collection of half-open intervals is known as the ``barcode'' \cite{ghrist2008} of $X$, and we use it to visualize the $m$th persistent homology. Generators with longer associated half-open intervals are more persistent. In general, one uses the persistence of features to distinguish signal from noise, but recent work indicates that persistence is not always readily interpretable in a meaningful way \cite{doi:10.1063/1.4978997,blumberg2018}. In our computations, we find that using traditional distance-based constructions on the \emph{LA Times} voting data yields ambiguous results about the persistence of features. However, by making an appropriate choice of construction for the persistence complex (see Subsections~\ref{ss:adj} and~\ref{ss:ls}), we can construct barcodes for the data in which persistence becomes a useful property for separating genuine features from noise.

We also draw attention to the distinction between the dimension of our embedding and the dimension of the topological object that we are studying, as we will be referencing both in the remainder of our paper. When we refer to the dimension of a simplicial complex, we mean the dimension of the highest-dimensional simplex in the simplicial complex. Similarly, the dimension of a homological feature refers to the dimension of the homology group in which it sits; that is, a feature in the $m$th persistent homology has dimension $m$. By contrast, when we refer to 2D data sets, we mean that the data are embedded in a two-dimensional ambient space. Therefore, a point set in 2D is a 0D object that lives in a 2D space. A feature in $H_1$ is a 1D feature that we can visualize as a loop that lives in 2D space. A precinct is a 2D object that is embedded in a 2D space using its latitude and longitude. In this paper, we refer to data sets and geographical maps based on the dimension of the space in which they live, whereas we refer to simplicial complexes, homology groups, and homological features based on the dimensionality of the objects themselves.

%%%%%

\section{Methods for Constructing Filtered Simplicial Complexes}
\label{sec:methods}

In this section, we describe the various methods that we use for constructing simplicial complexes from the voting data. The geographic data comes in the form of {\sc shapefiles}; it is a collection of polygons, rather than a point cloud. Although we do include computations based on existing constructions, which use point-cloud data, we also leverage the additional information inherent in the polygon form of geographical maps to suggest two new constructions that are better suited to our application. In Section~\ref{ss:sccomp}, we explain why one should expect these new approaches to yield better results. We perform computations using both these new constructions and two traditional ones, and we compare their performance in Section~\ref{sec:results}.

%%%%%

\subsection{Vietoris--Rips Complexes, Alpha Complexes, and Other Distance-Based Constructions}
\label{ss:vr}

We begin by reviewing several common methods for constructing filtered simplicial complexes from point clouds. One of the most prevalent constructions is the Vietoris--Rips (VR) complex, which one constructs using the pairwise distances\footnote{In the present paper, our measures of distance are metrics in the mathematical sense.} between points in a point cloud \cite{Vietoris1927,edelsbrunner2010}. 

Let $X$ be a data set in the form of a point cloud. Given a real number $\epsilon > 0$, we define the VR complex $\text{VR}_\epsilon(X)$ as follows:
\begin{equation*}
	\text{VR}_\epsilon(X) := \{\sigma \subset X: \,\, \forall \, x,y \in \sigma\,, \,\,d(x,y) \leq \epsilon\} \,.
\end{equation*}	
If there are $n$ points in $X$, the maximal possible VR complex is the $(n-1)$-simplex that consists of all points in $X$ and all of its subsimplices. By taking a collection $\{\epsilon_i\}$, with $0 = \epsilon_0 < \epsilon_1 < \epsilon_2 < \cdots < \epsilon_k$, and considering 
\begin{equation*}
	X = \text{VR}_{\epsilon_0}(X) \subseteq \text{VR}_{\epsilon_1}(X) \subseteq \cdots \subseteq \text{VR}_{\epsilon_k}(X) \,,
\end{equation*}	
we obtain a filtered simplicial complex whose PH we can compute. It is straightforward to construct a VR complex, because we only need to compute pairwise distances. Additionally, there are various fast algorithms for constructing it \cite{Zomorodian_2010}. Unfortunately, for large point clouds, the worst-case VR complex has $2^{|X|}-1$ simplices and dimension $|X|-1$. The largest county in our data set is Los Angeles County, which has almost $5000$ precincts, resulting in a worst-case complexity of about $2^{5000}-1$ simplices. This is very problematic.

The large number of precincts in several counties makes it intractable to compute VR complexes for these counties. For county--candidate combinations with at least $151$ precincts, we instead compute alpha complexes. The alpha complex \cite{1056714}, which we denote by $\text{A}_\epsilon(X)$, also relies on a distance parameter and is defined as follows. Let $\epsilon > 0$, and let $X_\epsilon := \bigcup_{x \in X} B(x,\epsilon)$. Additionally, let $(V_x)_{x \in X}$ be the Voronoi diagram of $X$. Consider the intersection $V_x \cap B(x,\epsilon)$ for each $x \in X$, and note that the collection of these sets covers $X_\epsilon$. We then have
\begin{equation*}
	\text{A}_\epsilon(X) = \{\sigma \subset X : \,\, \forall \, x_i \in \sigma\,, \,\,\bigcap_i \left(V_x \cap B(x,\epsilon)\right) \neq \emptyset\} \,.
\end{equation*}	
Because of the restriction of the $\epsilon$-balls to the Voronoi diagram, the alpha complex restricts the dimension of the space in which $X$ is embedded. In our case, because our data is embedded in $\mathbb{R}^2$, the alpha complex of a county has 2D simplices (e.g., faces) as its highest-dimensional simplices. 

The two constructions above both require the input data to be in the form of a point cloud. To each precinct, we associate its centroid, which we calculate according to (latitude, longitude) coordinates using the centroids plug-in in QGIS \cite{qgis}. We directly compute the Euclidean distance between the (latitude, longitude) coordinates of precinct centroids. Therefore, we do not make a choice of map projection.
 
Vietoris--Rips complexes are employed commonly, because it is relatively easy to construct them, they are intuitively appealing, they have important theoretical guarantees from the Nerve Theorem \cite{kerber2013}, and (perhaps most importantly) they have been implemented widely in existing PH packages. In general, $\epsilon$-ball thickenings are a natural way to approach the problem of approximating a space from which one has only a sample of points. Points that are close to each other should be much more likely to be connected in the space than points that are far apart from each other, and thickenings also capture the intuition of blurring an image by squinting at it until the points start to merge. However, for our purposes, the point clouds that we construct do not bear a strong visual resemblance to the geographical maps from which we construct them, and the locations of holes in these maps are independent of distance. In Figures~\ref{fig:3_vr} and~\ref{fig:3_alpha}, we show visualizations of VR and alpha complexes for one of our precincts. {Note that we consider only the red precincts (or, alternately, only the blue precincts); we make this simplification both to decrease computational complexity and to preserve an intuitive notion of closeness in voting patterns, as well as in geography. In these visualizations, observe that the complexes do not visually resemble the underlying geographical maps, and they also appear to have rather different topological properties. To address these issues, we propose two novel constructions in the next two subsections.

\begin{figure}[!htbp]
\begin{subfigure}[b]{.19\textwidth}
	\includegraphics[width=.95\textwidth]{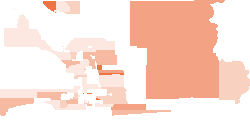}
	\subcaption{}
	\label{fig:3_vra}
\end{subfigure}
\begin{subfigure}[b]{0.2\textwidth}
	\includegraphics[width=.95\textwidth]{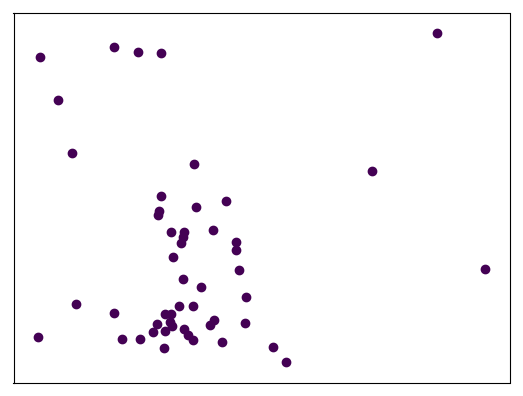}
	\subcaption{}
	\label{fig:3_vrb}
\end{subfigure}%
\begin{subfigure}[b]{0.2\textwidth}
	\includegraphics[width=.95\textwidth]{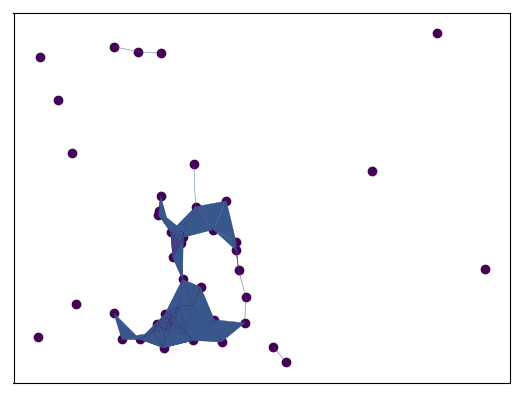}
	\subcaption{}
	\label{fig:3_vrc}
\end{subfigure}%
\begin{subfigure}[b]{0.2\textwidth}
	\includegraphics[width=.95\textwidth]{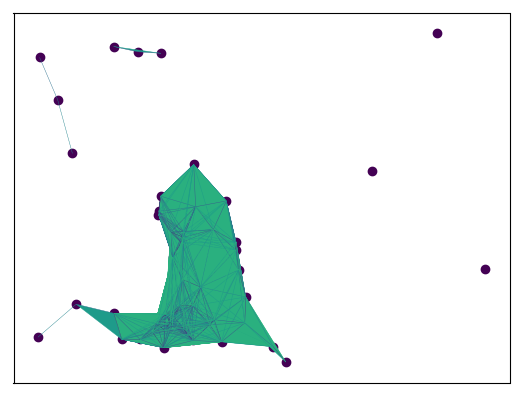}
	\subcaption{}
	\label{fig:3_vrd}
\end{subfigure}%
\begin{subfigure}[b]{0.2\textwidth}
	\includegraphics[width=.95\textwidth]{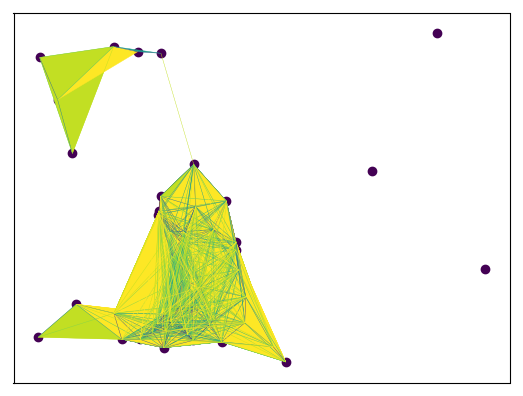}
	\subcaption{}
	\label{fig:3_vre}
\end{subfigure}%
\vspace{-.50cm}
\caption{Illustration of a Vietoris--Rips complex on the \emph{LA Times} voting data. (a) The red precincts (in which more people voted for Donald Trump than for Hillary Clinton) of Imperial County in 2016. In panels (b)--(e), we show the VR complex that approximates the county, with each successive image showing the VR complex as we increase $\epsilon$. Observe that the contiguous region in the east of the county is not captured by this complex and that the western region includes a large number of 1-simplices and 2-simplices, despite the fact that there are relatively few precincts on the map. Both phenomena occur because the eastern precincts are much larger, so their centroids are much farther apart than the small (but not necessarily contiguous) precincts in the west.
}
\label{fig:3_vr}
\end{figure}

\begin{figure}[!htbp]
\begin{subfigure}[b]{.19\textwidth}
	\includegraphics[width=.95\textwidth]{images/maps/025-imperial-trump}
	\subcaption{}
	\label{fig:3_alphaa}
\end{subfigure}
\begin{subfigure}[b]{0.2\textwidth}
	\includegraphics[width=.95\textwidth]{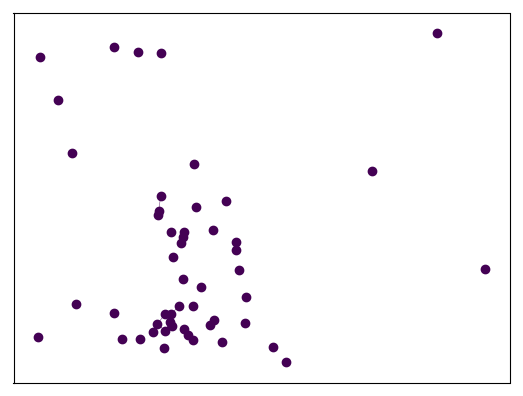}
	\subcaption{}
	\label{fig:3_alphab}
\end{subfigure}%
\begin{subfigure}[b]{0.2\textwidth}
	\includegraphics[width=.95\textwidth]{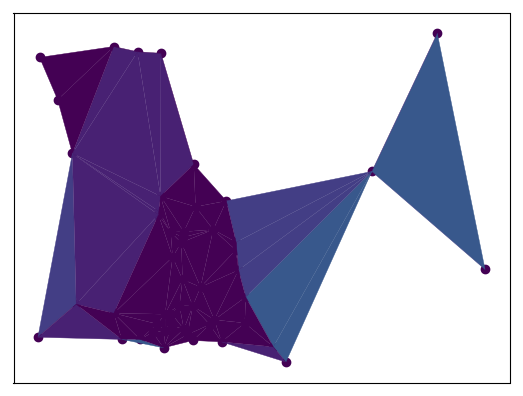}
	\subcaption{}
	\label{fig:3_alphac}
\end{subfigure}%
\begin{subfigure}[b]{0.2\textwidth}
	\includegraphics[width=.95\textwidth]{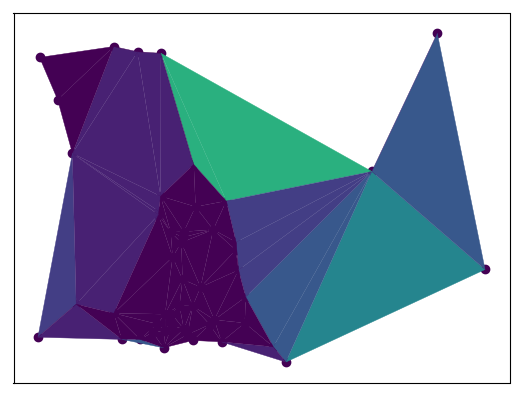}
	\subcaption{}
	\label{fig:3_alphad}
\end{subfigure}%
\begin{subfigure}[b]{0.2\textwidth}
	\includegraphics[width=.95\textwidth]{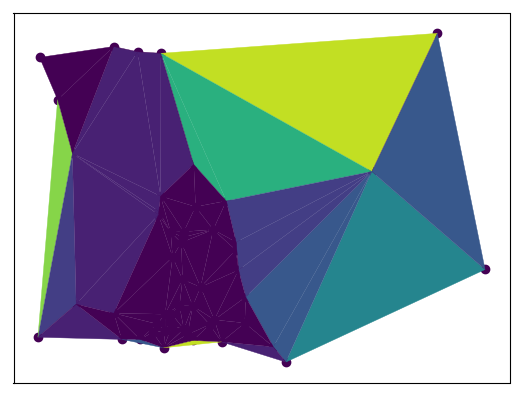}
	\subcaption{}
	\label{fig:3_alphae}
\end{subfigure}%
\vspace{-.50cm}
\caption{Illustration of an alpha complex on the \emph{LA Times} voting data. (a) The red precincts (in which more people voted for Donald Trump than for Hillary Clinton) of Imperial County in 2016. In panels (b)--(e), we show the alpha complex that approximates the county, with each successive image showing the complex as we increase $\epsilon$. Observe that the filtered simplicial complex has much larger 2-simplices than what we obtained for VR complexes (see Figure~\ref{fig:3_vr}), and that (unlike in Figure~\ref{fig:3_vr}) once the western region is covered by 2-simplices (which, as one can see in panel (c), occurs fairly early in the filtration), new 2-simplices do not arise as we increase $\epsilon$. However, similar to what we observed in the VR complex, the resulting simplicial complex yields a simply-connected region in the west; this does not accurately reflect the underlying geographical map.}
\label{fig:3_alpha}
\end{figure}

%%%%

\subsection{Adjacency Complexes}
\label{ss:adj}

Our first new type of construction of a filtered simplicial complex is based on the notion of a network adjacency. Consider a network whose nodes are precincts and whose edges are determined by ``queen adjacency''. We use the definition of queen adjacency from graphical information systems (GIS); two precincts are queen adjacent if they touch at any two points, including corners (which is a somewhat different idea from the movement of queens in chess). This is distinct from ``rook adjacency'', in which two precincts are adjacent if they share a boundary. Intuitively, we can imagine such a network as one in which any path in it corresponds to an ability to physically walk from one precinct to another in a contiguous fashion. Note that some precincts are not simply connected or even have multiple connected components. In Section~\ref{sec:results}, we discuss the effects of such features on our results.

By considering different levels of voter preferences for Donald Trump or Hillary Clinton, we construct a nested sequence of networks. We define a value
\begin{equation}
	\delta_{b,r}(p) = \frac{|V_b(p) - V_r(p)|}{|V_b(p) + V_r(p)|} \,,
\end{equation}	
where $V_b(p)$ is the number of blue (i.e., Clinton) votes in a precinct $p$ and $V_r(p)$ is the number of red (i.e., Trump) votes in that precinct. For example, for a given county, consider all of its precincts with a majority who voted for Hillary Clinton in 2016. For our first network, we consider only those precincts for which $\delta_{b,r}(p) \geq .95$. For the next network in the sequence, we take all precincts with $\delta_{b,r}(p) \geq .90$. We continue decreasing the strength of voter preference until we consider all precincts in which Clinton won, along with all of their adjacencies. At this stage, we stop and construct a filtered simplicial complex of dimension-$1$ simplicial complexes. To incorporate faces, we add a 2-simplex between any three nodes that are all pairwise adjacent. This gives a dimension-$2$ filtered simplicial complex, on which we can perform PH computations.

Using network adjacencies allows us to retain spatial information about our precincts that we lose when we consider only the point cloud of precinct centroids. In our application to voting data, our adjacency construction captures a notion of contiguity that is missing from the existing distance-based constructions. In Figure~\ref{fig:3_adj}, we show an example of a filtered simplicial complex, which we construct using adjacencies, that approximates Imperial County. It has better contiguity properties than the VR and alpha complexes that we showed in Figures~\ref{fig:3_vr} and~\ref{fig:3_alpha}. However, this adjacency approach still requires us to associate a single point to each precinct polygon, rather than considering the entire area that it covers. This suggests another possible construction (based on level sets), which we describe in Section~\ref{ss:ls}.

\begin{figure}[!htbp]
\begin{subfigure}[b]{.19\textwidth}
	\includegraphics[width=.95\textwidth]{images/maps/025-imperial-trump}
	\subcaption{}
	\label{fig:3_adja}
\end{subfigure}
\begin{subfigure}[b]{0.2\textwidth}
	\includegraphics[width=.95\textwidth]{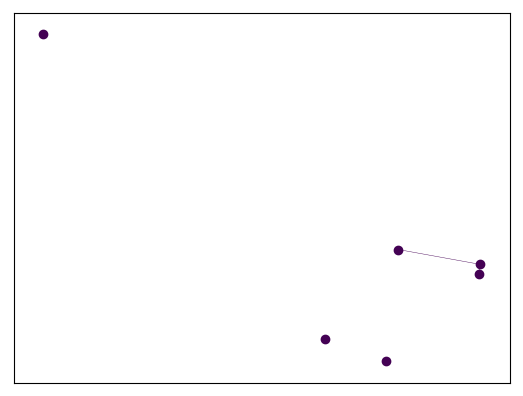}
	\subcaption{}
	\label{fig:3_adjb}
\end{subfigure}%
\begin{subfigure}[b]{0.2\textwidth}
	\includegraphics[width=.95\textwidth]{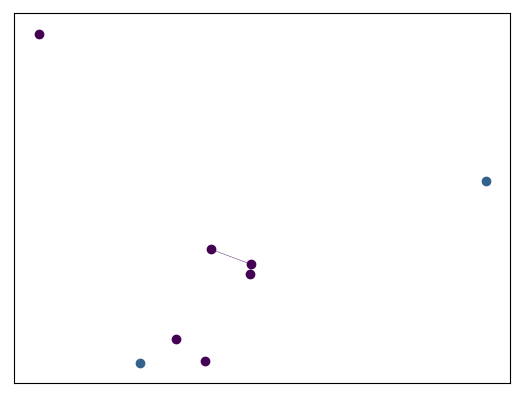}
	\subcaption{}
	\label{fig:3_adjc}
\end{subfigure}%
\begin{subfigure}[b]{0.2\textwidth}
	\includegraphics[width=.95\textwidth]{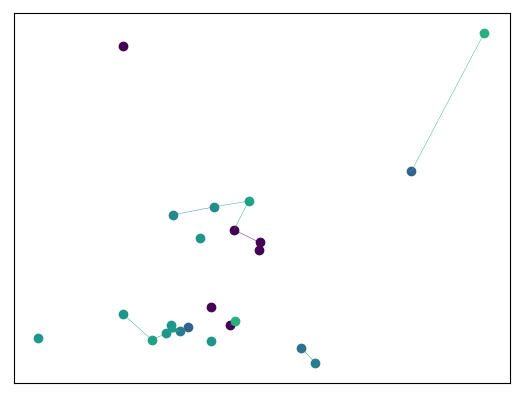}
	\subcaption{}
	\label{fig:3_adjd}
\end{subfigure}%
\begin{subfigure}[b]{0.2\textwidth}
	\includegraphics[width=.95\textwidth]{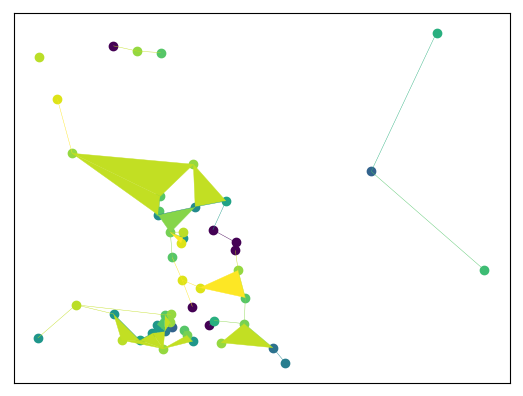}
	\subcaption{}
	\label{fig:3_adje}
\end{subfigure}%
\vspace{-.50cm}
\caption{Illustration of an adjacency complex on the \emph{LA Times} voting data. (a) The red precincts (in which more people voted for Donald Trump than for Hillary Clinton) of Imperial County in 2016. In panels (b)--(e), we show an associated adjacency complex that approximates the county, where we order the panels based on decreasing strength of preference for Trump. In panel (e), we observe that the eastern region is simply connected and that the western region has many 1-simplices. However, the latter is not covered by 2-simplices, so it is not simply connected. Although the depicted filtered simplicial complex does not seem to closely resemble the geographical map in Figure~\ref{fig:3_adja} visually, its topological properties do appear to be similar.
}
\label{fig:3_adj}
\end{figure}

\vspace{-.75cm}

%%%%%

\subsection{Level-Set Complexes}
\label{ss:ls}

The second new method that we introduce is one that leverages the manifold nature of our data. For the previous methods (namely, the VR, alpha, and adjacency complexes), we were forced to make choices in how to assign precincts to points. For the VR and alpha constructions (i.e., the distance-based ones), we also had to make a choice of embedding into Euclidean space. We now introduce a complex based on level sets; for it, we use polygon {\sc shapefiles} as input and evolve them using level-set equations for motion of interfaces. In this section, we give an overview of the filtered simplicial complex that we generate using the level-set method. The level-set method was introduced in \cite{OSHER198812}; we offer an intuitive explanation in this section, and further details are available at \cite{osher2003}.

Let $M$ denote the 2D manifold that consists of the collection of all of a county's precincts that voted for the same candidate (regardless of the strength of the majority). We construct a sequence of manifolds,
\begin{equation*}
	M = M_0 \subset M_1 \subset \cdots \subset M_n \,,
\end{equation*}	
by considering the boundary $\Gamma$ of $M$ and performing front propagation on it so that the boundary expands outward, resulting in a larger manifold. We use the level-set method to efficiently solve the front-propagation problem. To do this, we evolve a function $\phi(\vec{x},t): \mathbb{R}^2 \times \mathbb{R} \to \mathbb{R}$ according to the level-set equation
\begin{equation} \label{levelset}
	\frac{\partial \phi}{\partial t} = v \left|\nabla \phi\right| \,,
\end{equation}
where $v$ is velocity.

By assigning the initial condition $\phi(\vec{x},0)$ to be the signed distance function from $\vec{x}$ to $\Gamma$, we see that the $0$-level set of $\phi(\vec{x},0)$ is precisely the set of points that lie on $\Gamma$. When we evolve $\phi$ according to \eqref{levelset} up to time $T$, the resulting $0$-level set of $\phi(\vec{x},T)$ gives $\Gamma_T$, the expansion of $\Gamma$ that results from movement normal to the boundary at velocity $v$.

Intuitively, in terms of our geographical map, we can visualize the graph of $\phi(\vec{x},0)$ as a mountain (or multiple mountains, if there is more than one connected component), with the boundary of the map at sea level, the interior of the map above water, and the complement of the map below water. The set $M_0$ is the set of points $\vec{x}$ that are at or above sea level. As we evolve $\phi$, we move the entire mountain upward, increasing the amount of land above water. The new region that is at or above water is our expanded manifold $M_T$. In Figure~\ref{fig:3_sanmateo_ls_3d}, we show the evolution of the $0$-superlevel set (i.e., all points $\vec{x}$ such that $\phi(\vec{x},t)\geq0$) as $T$ increases, along with the graph of $\phi$ to help visualize the corresponding evolution of the level-set equation~\eqref{levelset}.

\begin{figure}[!htbp]
\centering
\begin{subfigure}{0.33\textwidth}
\centering
\includegraphics[width=.95\textwidth, trim = {3.6cm 1.4cm 3.2cm 1.4cm},
clip]{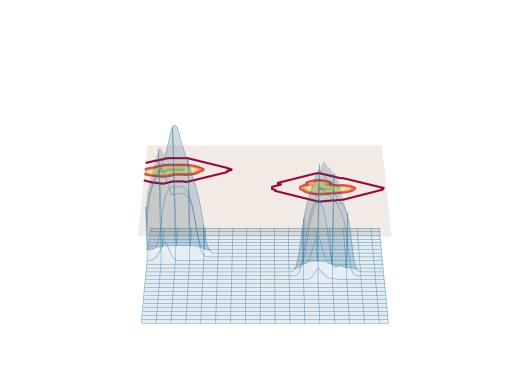}
\includegraphics[width=.95\textwidth, trim = {0 3cm 0 2cm},
clip]{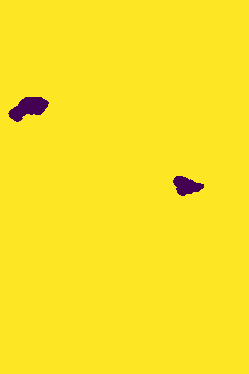}
\subcaption{$T=0$}
\end{subfigure}%
\begin{subfigure}{0.33\textwidth}
\centering
\includegraphics[width=.95\textwidth, trim = {3.6cm 1.4cm 3.2cm 1.4cm},
clip]{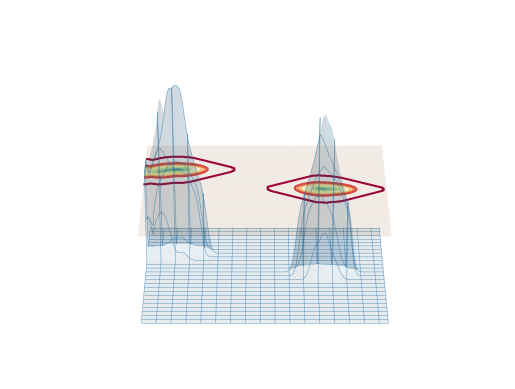}
\includegraphics[width=.95\textwidth, trim = {0 3cm 0 2cm},
clip]{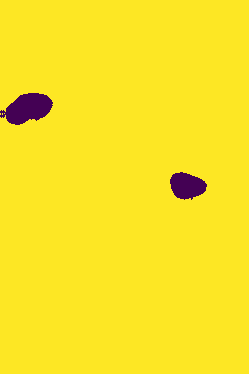}
\subcaption{$T=4$}
\end{subfigure}%
\begin{subfigure}{0.33\textwidth}
\centering
\includegraphics[width=.95\textwidth, trim = {3.6cm 1.4cm 3.2cm 1.4cm},
clip]{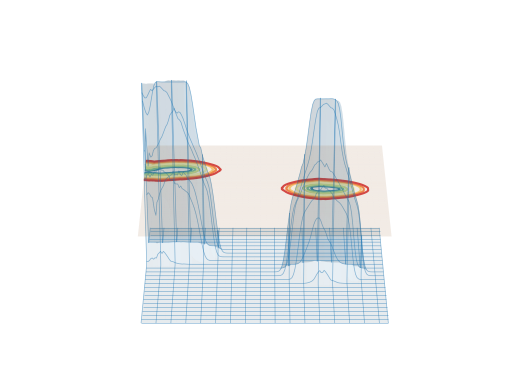}
\includegraphics[width=.95\textwidth, trim = {0 3cm 0 2cm},
clip]{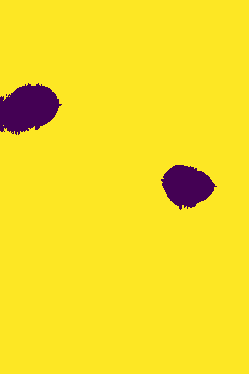}
\subcaption{$T=10$}
\end{subfigure}%
\vspace{-.5cm}
\caption{Evolution of (top row) a level set on red precincts (in which more people voted for Donald Trump than for Hillary Clinton) in San Mateo County, with corresponding (bottom row) contour plots of $\phi$. As $T$ increases, the graph of $\phi$ translates upward, so that the $0$-superlevel set expands. (Clipping of minimum and maximum values, which we do for computational efficiency, leads to flat areas at the minimum and maximum values of $\phi$.)
}
\label{fig:3_sanmateo_ls_3d}
\end{figure}

In Figure~\ref{fig:3_sanmateo_ls}, we show another such sequence of manifolds, which we obtain by evolving a level set on blue precincts in San Mateo County. The original geographical map has holes of various sizes, and the amount of time that it takes for a given hole to disappear is longer for smaller holes.

\begin{figure}[!htbp]
\begin{subfigure}{0.2\textwidth}
	\includegraphics[width=.95\textwidth]{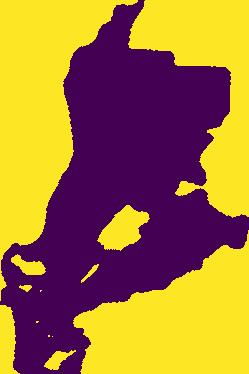}
\end{subfigure}%
\begin{subfigure}{0.2\textwidth}
	\includegraphics[width=.95\textwidth]{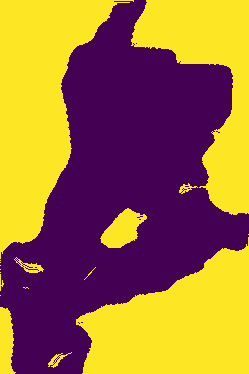}
\end{subfigure}%
\begin{subfigure}{0.2\textwidth}
	\includegraphics[width=.95\textwidth]{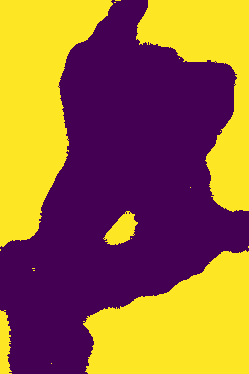}
\end{subfigure}%
\begin{subfigure}{0.2\textwidth}
	\includegraphics[width=.95\textwidth]{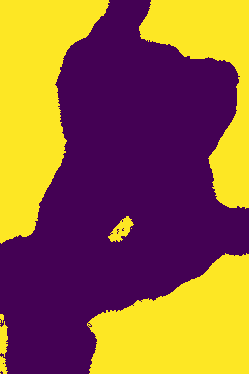}
\end{subfigure}%
\begin{subfigure}{0.2\textwidth}
	\includegraphics[width=.95\textwidth]{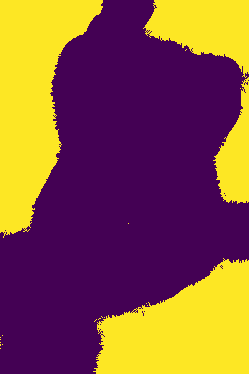}
\end{subfigure}%
\caption{Evolution of a level set on blue precincts (in which more people voted for Hillary Clinton than for Donald Trump) in San Mateo County.}
\label{fig:3_sanmateo_ls}
\end{figure}

To turn this sequence into a filtered simplicial complex, we choose a triangulation of the plane and impose each $M_T$ over this triangulation in the following manner. In our triangulation, (1) every 5th pixel is a vertex and (2) each vertex is connected to its four neighbors in the cardinal directions, as well as to its northwest and southeast neighbors. Other triangulation choices are also viable, but ours is computationally convenient (because it limits the number of vertices) and is easy to visualize. 
We use the rule that if all vertices of a 2-simplex lie within $M_T$, we add that simplex and all subsimplices to the corresponding simplicial complex $X_T$. This yields the filtered simplicial complex
\begin{equation*}
	X_0 \subset X_1 \subset \cdots \subset X_n \,.
\end{equation*} 
We evolve until a time $T$ that is sufficiently large for all holes to close. (The geographical maps are in a bounded subset of $\mathbb{R}^2$, so such a time is guaranteed to exist.) For more implementation details, see Appendix~\ref{app:algls}.

The greatest strength of our level-set approach to constructing a simplicial complex is that it gives an explicit triangulation of a geographical map that does not depend on any choice assigning precincts to points. The simplicial complexes that we build using the level-set method thus embed nicely into the plane, and they more closely resemble the underlying geographical maps from which we start than the other examined methods. Moreover, persistence is nicely interpretable for the level-set approach. Any hole that exists in the geographical map also exists in the initial complex (as long as the hole is not finer than the employed triangulation of $\mathbb{R}^2$), so every hole is a feature that is born at time $0$. The persistence of the feature then indicates the distance scale on which it exists. In this manner, we can distinguish between short-persistence true features and short-persistence noise due to the evolution, because short-persistent noise does not appear until later time steps in the level-set evolution. (An example of this occurs in Figure~\ref{fig:3_sanmateo_ls}, where a bay on the eastern side of the map is not a closed loop in the leftmost image, but it is closed in the next image because the opening of the bay is smaller than the bay itself.) Furthermore, although the level-set complex still suffers from the sensitivity to scale of other distance-based constructions, it does not require us to make a scaling choice, as is necessary for existing distance-based constructions. Both very large and very small holes are captured immediately, because the connectedness of a simplex does not rely on the distance between precinct centroids. In Figure~\ref{fig:3_ls}, we show an example of a level-set simplicial complex for a voting map of Imperial County.

\begin{figure}[!htbp]
\begin{subfigure}[b]{.19\textwidth}
	\includegraphics[width=.95\textwidth]{images/maps/025-imperial-trump}
	\subcaption{}
	\label{fig:3_lsa}
\end{subfigure}
\begin{subfigure}[b]{0.2\textwidth}
	\includegraphics[width=.95\textwidth]{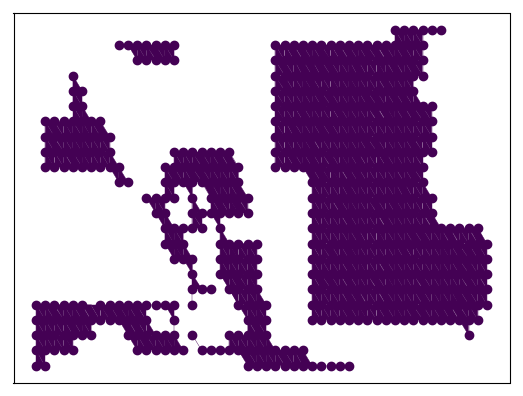}
	\subcaption{}
	\label{fig:3_lsb}
\end{subfigure}%
\begin{subfigure}[b]{0.2\textwidth}
	\includegraphics[width=.95\textwidth]{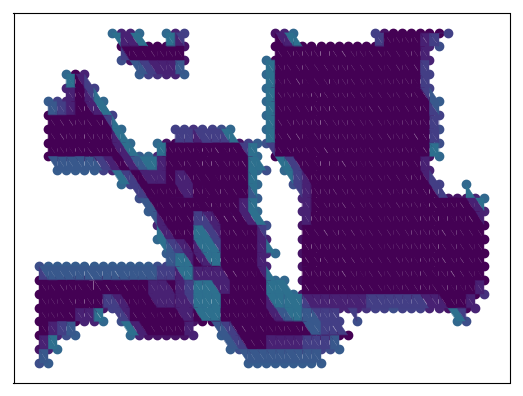}
	\subcaption{}
	\label{fig:3_lsc}
\end{subfigure}%
\begin{subfigure}[b]{0.2\textwidth}
	\includegraphics[width=.95\textwidth]{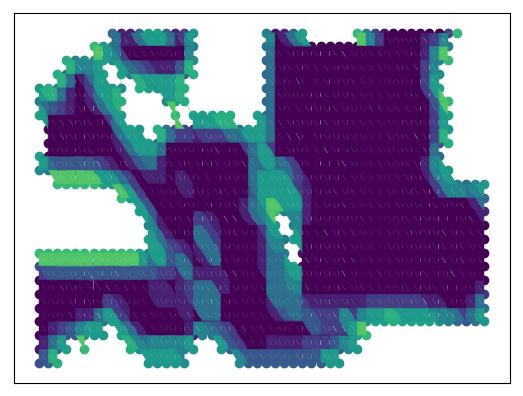}
	\subcaption{}
	\label{fig:3_lsd}
\end{subfigure}%
\begin{subfigure}[b]{0.2\textwidth}
	\includegraphics[width=.95\textwidth]{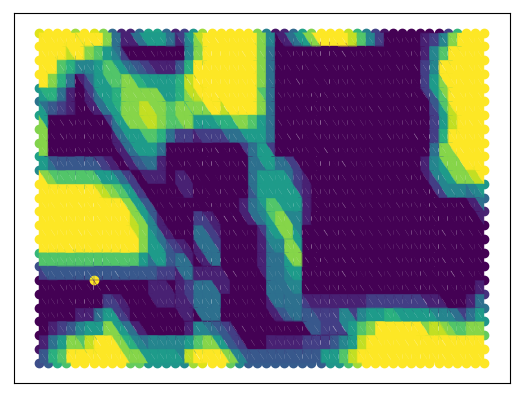}
	\subcaption{}
	\label{fig:3_lse}
\end{subfigure}%
\vspace{-.5cm}
\caption{Illustration of a level-set complex on the \emph{LA Times} voting data. (a) The red precincts of Imperial County in 2016. In panels (b)--(e), we show the level-set complex that is associated with the voting map of Imperial County. We order it according to the number of time steps in the level-set evolution. Observe in panel (b) that the complex immediately resembles the original voting map and that the smaller holes are filled in faster than the large ones. Given enough time steps, the level set will evolve to cover the entire bounding box that we show in the figure.
}
\label{fig:3_ls}
\end{figure}

\vspace{-.75cm}

%%%%%%

\subsection{Comparing the Simplicial-Complex Constructions}
\label{ss:sccomp}

In the previous subsections, we briefly discussed some of the ideas that the different constructions of simplicial complexes are intended to capture. We now give more detail why these ideas are particularly useful for applications to geospatial data. Persistent homology on point clouds is based largely on the idea that the distance between points indicates something meaningful about their similarity or connectedness. Under this assumption, points that are close together have a fundamentally different relationship to each other than points that are far apart. Consequently, features that occur at small distance scales should not represent the same behaviors as features that occur at large ones. However, in our case (and in other applications to geospatial data), we observe two problems: (1) districts with isolated voting patterns occur at a variety of distance scales; and (2) physical distance does not correspond to geographic connectedness. More generally, spatial applications for which the information of interest is not encoded in distances may suffer from both issues. We refer to the first issue as ``scaling'' and to the second issue as ``contiguity''. 

In the next two subsubsections, we discuss why existing PH constructions struggle with scaling and contiguity, as well as which of our methods address them and how. In Table~\ref{tab:3_comp}, we summarize the methods and their performance. One potential solution to the problem of physical distance being unrepresentative is to replace it with some other distance and to perform PH using the new distance as the filtration parameter. Unfortunately, this is an undesirable solution for many applications to spatial data: although Euclidean distance between points may not encode the features of interest, the embeddedness of the data into space is often relevant; changing the notion of distance may not reflect this fact, or it may force one to make a choice for how to combine multiple notions of distance. By contrast, our new methods for computing PH allow us to incorporate the spatial embedding of data without reducing that embedding to a set of pairwise distances between points, while also potentially avoiding the scaling and contiguity issues that arise from distance-based constructions.

\begin{table}[!htbp]
\centering
\caption{Comparison of various methods of constructing simplicial complexes, based on whether they address scaling and contiguity problems.}
\label{tab:3_comp}
\begin{tabular}{l c c c c}
\toprule
Issue & VR & Alpha & Adjacency & Level set \\
\midrule
Scaling & \xmark & \xmark & \checkmark & \xmark \\
Contiguity & \xmark & \xmark & \checkmark & \checkmark \\
\bottomrule
\end{tabular}
\end{table}

%%%%

\subsubsection{Scaling}

When associating precincts to point clouds, the physical distance between precincts is based mostly on the extent to which the area is urban or rural. Accordingly, distance constructions result in very few persistent features. For the most part, rural areas are not connected enough at small scales to generate interesting features, whereas urban areas are too connected at large scales to preserve features. This implies that many meaningful features (e.g., a single red pocket in an urban community) are not persistent. Even worse, the most persistent features give information about whether there are densely populated areas that surround relatively unpopulated ones, but they give little meaningful information about the underlying political inclinations of those regions. These results counter the conventional wisdom about PH that the strongest signals should come from the most persistent features and that non-persistent features are likely to be the result of noise. 

This leaves us with two possibilities: either (1) we evaluate the features that result from PH using criteria that do not depend solely on examining the most persistent features; or (2) we must find other ways of constructing simplicial complexes, such that persistence becomes a meaningful quantity to compute for the problem of interest. There exists work on the former approach \cite{Bubenik:2015:STD:2789272.2789275, Adams:2017:PIS:3122009.3122017, bobrowski2017, NIPS2015_5887, Reininghaus2015ASM, Zhu:2016:SMP:3060832.3060964}, and our work complements this prior research by adopting the latter approach. In our adjacency method, by letting the filtration parameter be strength of voting preference rather than distance, we are able to interpret persistence as a measure of the difference between the preferences of a ``hole'' compared to preferences of the areas that surround it. That is, more persistent features represent holes with voting results that are very different from their neighbors. Consequently, the most persistent features are exactly the most meaningful ones, as they indicate which regions have the strongest outlying signals.

%%%%%

\subsubsection{Contiguity} 

For our PH computations to be meaningful, we want the simplicial complexes that we build to approximate our data as closely as possible. For the VR and alpha constructions, we assumed that precincts (i.e., points) are connected as long as their centroids are close enough. In practice, whether two precincts are adjacent has little to do with the distance between them. In urban areas, precincts that are very close to each other may have other precincts sandwiched between them, such that they are not connected. In rural ones, by contrast, precincts whose centroids are very far apart from each other may in fact be contiguous. Both the adjacency and level-set constructions address this issue.

In the adjacency approach, we define the adjacency matrix of a network based on whether two precincts share a border. As a result, all of the 1-simplices in our filtered simplicial complex come directly from physical contiguity. In the level-set approach, because our input data comes in the form of a manifold, both the 1-simplices and 2-simplices reflect the physical contiguity of the original geographical maps. Both approaches allow us to construct simplicial complexes that seem to better approximate the data than the traditional distance-based approaches. See our illustration in Figure~\ref{fig:3_cont}. Note that it may be possible to improve a distance-based construction by using the minimum distance between points in a precinct, rather than the distance between centroids (or between other representative points). However, this is a sufficiently difficult computational problem that we do not expect it to be a practical solution.

\begin{figure}[!htbp]
\centering
\begin{subfigure}{0.33\textwidth}
	\centering\includegraphics[width=\textwidth]{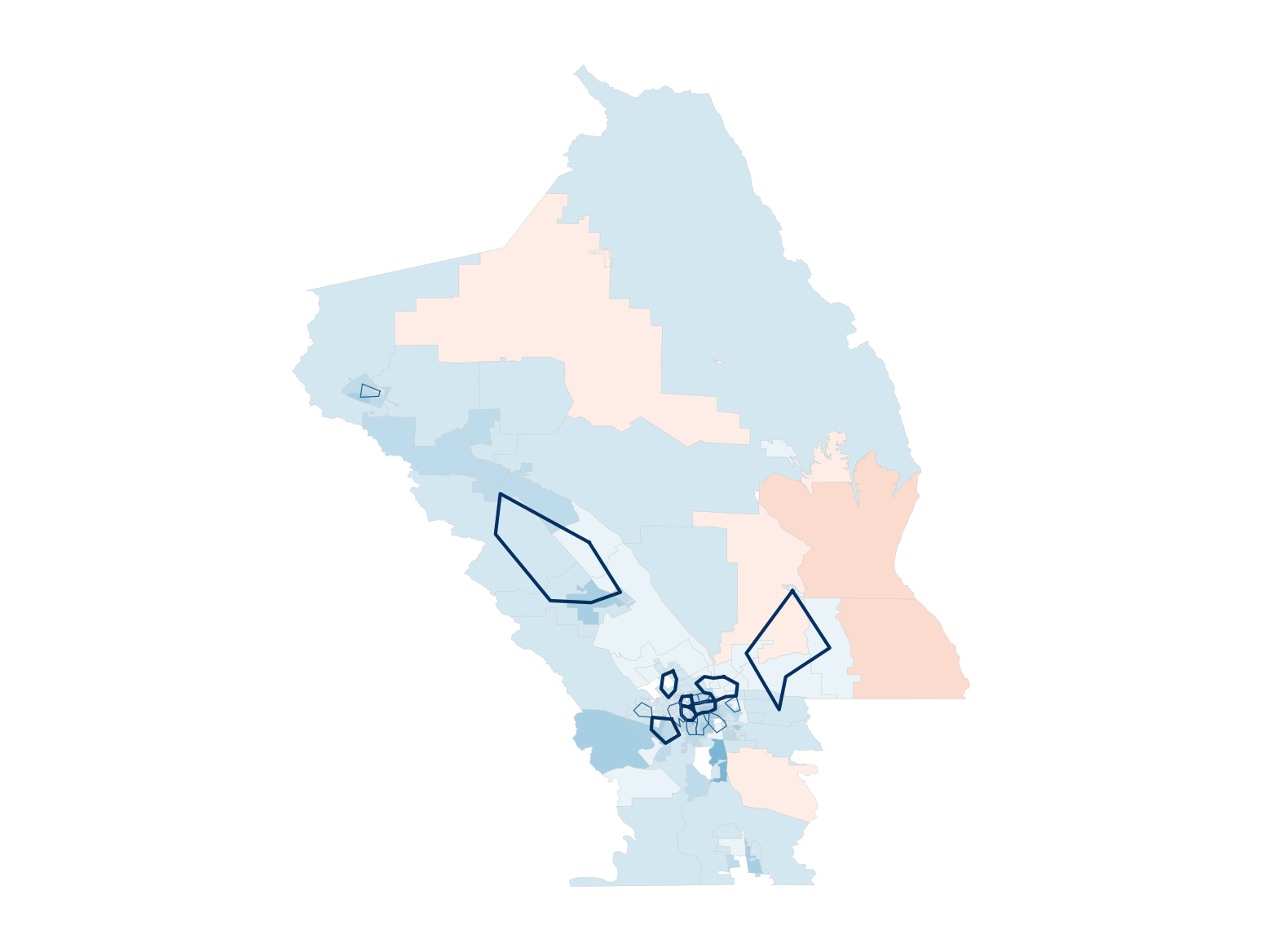}
\vspace{-.5cm}
	\caption{\label{fig:3_cont_a} Vietoris--Rips complex}
\end{subfigure}%
\begin{subfigure}{0.33\textwidth}
	\centering\includegraphics[width=\textwidth]{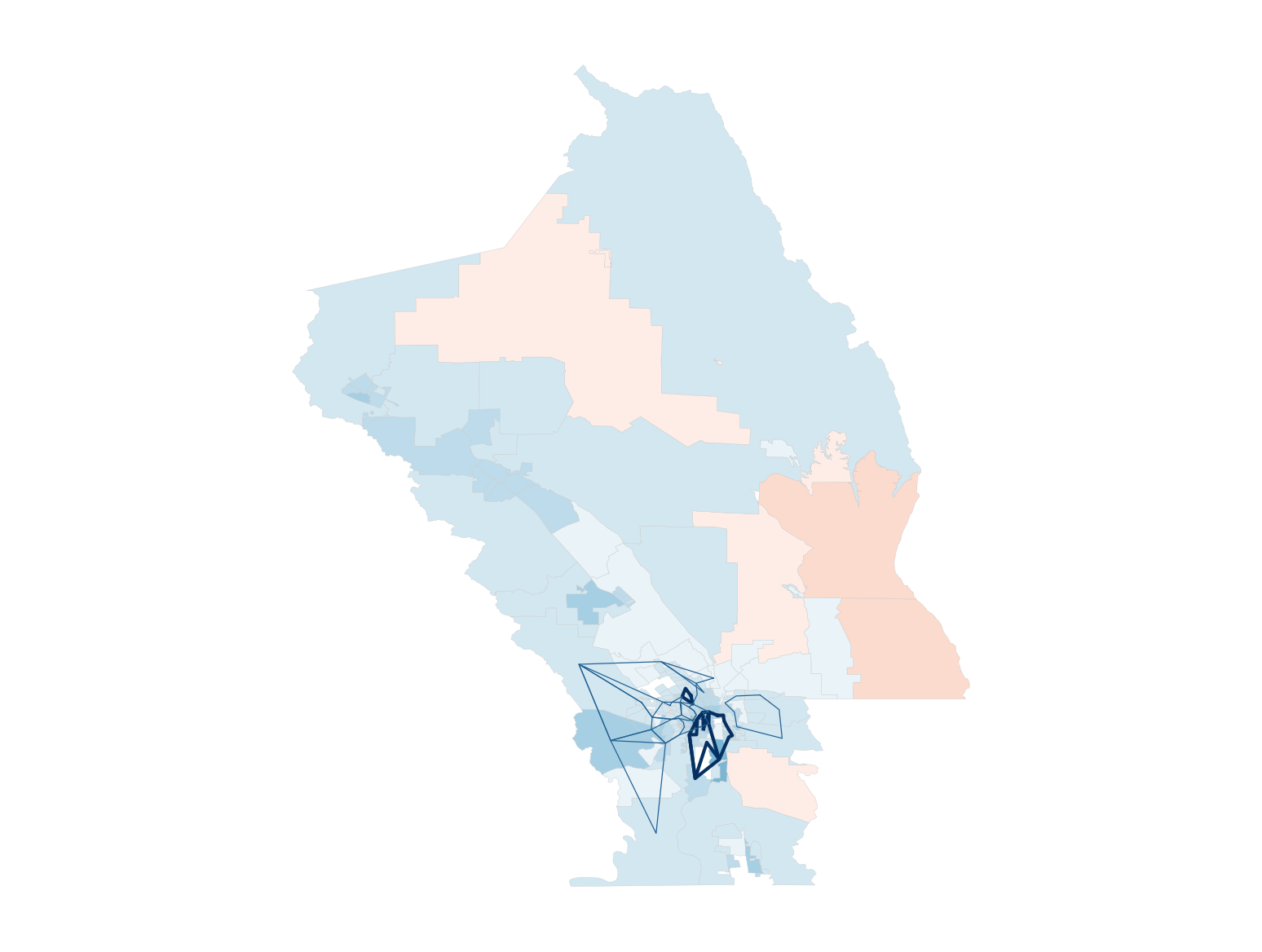}
\vspace{-.5cm}
	\caption{\label{fig:3_cont_b} Adjacency complex}
\end{subfigure}
\vspace{-.5cm}
\caption{Napa County, with the generators of features in $H_1$ marked as cycles in dark blue. In (a), we observe at least one ``loop'' in the eastern part of Napa County that is not contiguous, as it is composed of several small precincts whose union is not connected. In (b), by contrast, the adjacency construction captures several loops, each of which has generators whose union forms a contiguous region.}
\label{fig:3_cont}
\end{figure}

\vspace{-.75cm}

%%%%%

\section{Computational Results}
\label{sec:results}

In this section, we summarize our computational results. For the construction of VR and alpha complexes, we use the {\sc Python} package {\sc Gudhi} \cite{gudhi:urm,gudhi:RipsComplex, gudhi:AlphaComplex,gudhi:cython}. For the computation of PH and its generators, we use a modified version of {\sc Phat} \cite{10.1007/978-3-662-44199-2_24}, a \verb!C++! package for the fast computation of barcodes. We implement the adjacency and level-set constructions using an adaptation of the fast incremental VR algorithm \cite{Zomorodian_2010}. For details about our implementation and links to code, see Appendix~\ref{app:alg}.

%%%%

\subsection{Sizes and Computation Times}
\label{ss:resultssc}

Construction of simplicial complexes can be very slow, as one must check all possible simplices. The number of simplices grows as $n^d$, where $n$ is the number of vertices and $d$ is the maximum simplex dimension that one is considering. Consequently, methods that build smaller simplicial complexes tend to be faster. In Table~\ref{tab:4_size}, we compare the number of simplices in the simplicial complexes that we construct using the various methods. To keep computation times tractable, we compute VR complexes only for counties with at most $150$ precincts that voted for a certain candidate. If $151$ or more precincts voted for the same candidate, we instead compute alpha complexes.

From Table~\ref{tab:4_size}, we see that the adjacency and level-set complexes do not scale in size as rapidly as the VR complexes. This arises from how we construct these complexes. In adjacency complexes, the number of neighbors tends to be almost constant for any number of precincts, as realistically there are practical bounds on the number of precincts that can border another precinct. The beneficent scaling of the level-set complexes with respect to the number of precincts arises from our specific choices for how we construct them. Because we take each vertex of a simplicial complex to be a point on a triangular grid, it has at most six neighboring vertices (one for each of its cardinal directions, as well as one to its upper left and one to its lower right), and it can thus be a member of at most six 2-simplices. One can make different choices of triangular grids---in our case, we simply added a northwest/southeast diagonal to each square in a square grid---and the number of neighbors is $O(1)$, as long as the grid is composed of triangles that are roughly the same size and shape (as is true for many grids). However, even when the number of precincts is rather small, a level-set complex can still be rather large. Even when there are relatively few precincts, if those precincts constitute a large enough portion of a voting map, they will include many grid points and hence many vertices. In practice, we obtain a relatively large number of the possible 2-simplices in our level-set complexes, because our voting maps have large contiguous regions.

\begin{table}[!htbp]
\tiny
\centering
\caption{Sizes (i.e., number of simplices) of simplicial complexes. We first partition each county into precincts that voted for Clinton (C) and precincts that voted for Trump (T). We do not consider precincts that did not favor one of the two candidates. We then compute VR (or alpha), adjacency, and level-set complexes for each of these sets of precincts. (We compute VR complexes for counties with at most $150$ precincts and alpha complexes for counties with $151$ or more precincts.)
}
\label{tab:4_size}
\resizebox{13cm}{!} {
\begin{tabular}{l l l l l l l l l l}
\toprule
\multirow{2}{*}{County} & \# Precincts & \multicolumn{2}{c}{VR} & \multicolumn{2}{c}{Alpha} & \multicolumn{2}{c}{Adjacency} & \multicolumn{2}{c}{Level-set} \\
\cmidrule(lr){3-4}
\cmidrule(lr){5-6}
\cmidrule(lr){7-8}
\cmidrule(lr){9-10}

& & C & T & C & T & C & T & C & T \\
\midrule

Alameda & 1156 & -- & 1967 & 5843 & -- & 5755 & 70 & 3327 & 3578 \\
Alpine & 5 & 2 & 1 & -- & -- & 11 & 1 & 11962 & 1505 \\
Amador & 30 & 3 & 884 & -- & -- & 2 & 168 & 46 & 3979 \\
Calaveras & 29 & 8 & 641 & -- & -- & 6 & 92 & 1897 & 5195 \\
Colusa & 17 & 19 & 74 & -- & -- & 10 & 46 & 1665 & 5329 \\
Contra Costa & 711 & -- & 3551 & 3561 & -- & 3240 & 126 & 4135 & 3215 \\
Del Norte & 18 & 5 & 204 & -- & -- & 4 & 61 & 3584 & 6385 \\
El Dorado & 196 & 2397 & 89301 & -- & -- & 136 & 1123 & 782 & 4965 \\
Fresno & 592 & -- & -- & 1825 & 1431 & 1540 & 1192 & 2031 & 4788 \\
Glenn & 34 & 8 & 1152 & -- & -- & 4 & 156 & 329 & 5247 \\
Humboldt & 127 & 45998 & 680 & -- & -- & 504 & 119 & 15211 & 7323 \\
Imperial & 179 & 32496 & 6320 & -- & -- & 313 & 129 & 4375 & 6223 \\
Inyo & 25 & 33 & 216 & -- & -- & 14 & 51 & 4169 & 2242 \\
Kern & 642 & -- & -- & 1125 & 2119 & 928 & 2083 & 1429 & 5033 \\
Kings & 183 & 6305 & 69786 & -- & -- & 155 & 599 & 4849 & 7338 \\
Lake & 70 & 2279 & 779 & -- & -- & 99 & 73 & 4468 & 11275 \\
Lassen & 51 & 1 & 5920 & -- & -- & 1 & 250 & 193 & 11439 \\
Los Angeles & 4988 & -- & -- & 26551 & 1747 & 27705 & 1067 & 8587 & 6686 \\
Madera & 67 & 927 & 1947 & -- & -- & 103 & 132 & 925 & 5139 \\
Marin & 182 & -- & 3 & 1037 & -- & 1074 & 3 & 7893 & 621 \\
Mariposa & 25 & 5 & 401 & -- & -- & 7 & 91 & 2241 & 4485 \\
Mendocino & 250 & -- & 692 & 1115 & -- & 946 & 51 & 11901 & 1400 \\
Merced & 268 & 139832 & 54664 & -- & -- & 546 & 435 & 2213 & 6999 \\
Modoc & 21 & 0 & 399 & -- & -- & 0  & 94 & 0 & 7995 \\
Mono & 12 & 41 & 5 & -- & -- & 35 & 4 & 2499 & 3452 \\
Monterey & 467 & -- & 13887 & 2297 & -- & 1059 & 135 & 3597 & 4370 \\
Napa & 170 & 170093 & 56 & -- & -- & 858 & 15 & 10414 & 4968 \\
Nevada & 82 & 2569 & 2242 & -- & -- & 230 & 201 & 2946 & 2495 \\
Orange & 1668 & -- & -- & 5391 & 3811 & 4373 & 2632 & 5719 & 6513 \\
Placer & 363 & 5085 & -- & -- & 1685 & 141 & 1902 & 1210 & 3354 \\
Plumas & 30 & 8 & 618 & -- & -- & 6 & 102 & 723 & 6609 \\
Riverside & 1126 & -- & -- & 2291 & 2833 & 1602 & 2081 & 2231 & 2617 \\
Sacramento & 1267 & -- & -- & 2935 & 1275 & 15893 & 3459 & 4263 & 6748 \\
San Benito & 54 & 1804 & 276 & -- & -- & 152 & 67 & 699 & 6357 \\
San Bernardino & 2654 & -- & -- & 6206 & 4953 & 3658 & 2465 & 1700 & 6487 \\
San Diego & 2111 & -- & -- & 8007 & 3329 & 7480 & 2977 & 4680 & 7447 \\
San Francisco & 599 & -- & 0 & 3499 & -- & 3728 & 0 & 6826 & 0 \\
San Joaquin & 500 & -- & -- & 1659 & 1091 & 1490 & 902 & 7115 & 13419 \\
San Luis Obispo & 161 & 24600 & 14301 & -- & -- & 307 & 351 & 1319 & 4321 \\
San Mateo & 467 & -- & 8 & 2573 & -- & 2457 & 4 & 13865 & 782 \\
Santa Barbara & 250 & -- & 11950 & 971 & -- & 835 & 287 & 3488 & 6542 \\
Santa Cruz & 267 & -- & 28 & 1307 & -- & 1301 & 7 & 4737 & 295 \\
Shasta & 121 & 3 & 75177 & -- & -- & 2 & 745 & 941 & 5973 \\
Sierra & 22 & 3 & 233 & -- & -- & 2 & 57 & 417 & 3677 \\
Solano & 258 & 125438 & 13096 & -- & -- & 727 & 338 & 4589 & 5891 \\
Sonoma & 491 & -- & 886 & 2355 & -- & 2204 & 32 & 6031 & 899 \\
Stanislaus & 218 & 45984 & 51289 & -- & -- & 420 & 493 & 2536 & 6219 \\
Sutter & 52 & 62 & 3558 & -- & -- & 23 & 266 & 588 & 10689 \\
Tehama & 46 &0 & 4261 & -- & -- & 0 & 241 & 0 & 5007 \\
Trinity & 25 & 25 & 243 & -- & -- & 12 & 60 & 5485 & 10344 \\
Tulare & 250 & 13096 & -- & -- & 921 & 235 & 1032 & 2242 & 7763 \\
Tuolomne & 68 & 18 & 10605 & -- & -- & 6 & 334 & 3380 & 3997 \\
Yolo & 129 & 49597 & 486 & -- & -- & 559 & 70 & 5089 & 4597 \\
Yuba & 46 & 5 & 3422 & -- & -- & 3 & 199 & 1909 & 8521 \\
\bottomrule
\end{tabular}
}
\end{table}

In Table~\ref{tab:4_runtimes}, we compare the computation times for the construction and computation of PH for several of the larger (and therefore more computationally intensive) complexes. For a complete table of all computation times, see Appendix~\ref{app:tab}. From Table~\ref{tab:4_runtimes}, we see that our constructions of the adjacency and level-set complexes are significantly faster than construction of VR complexes, even for relatively small counties like El Dorado (which has only $196$ precincts). This is especially striking in light of the fact that we have not optimized our implementations of the new methods to make them as fast as possible. (For the level-set complexes, it is certainly possible to make the computations much faster using existing implementations of level-set dynamics \cite{GIBOU201882}.) 

We also see that our computations are only slightly slower than or of similar computation time to the construction of alpha complexes. Importantly, these speed gains are due largely to the significantly smaller number of simplices that we need for our new types of complexes. In 2D geospatial applications, the number of simplices is smaller than for other applications because of constraints from our starting geographical maps. In other applications, one does not typically benefit from such a built-in limitation in numbers. (For example, networks in general do not satisfy the property that the degrees of the vertices are roughly constant for any total number of vertices \cite{newman2018}.) However, other spatial applications (e.g., granular materials, transportation networks, and various examples in biology) will likely also benefit from these ideas.

\begin{table}[!htbp]
\centering
\caption{Computation times of selected county--candidate pairs, where we show the fastest method for each example in bold. We present several larger counties to show that our methods are substantially faster than computing VR complexes. For small counties, such as Imperial and Tulare, the improvement in computation time is less noticeable. Computing level-set complexes is not substantially faster for small counties than for large counties, as the number of simplices in a level-set complex is based on the resolution of the geographical map, rather than on the number of precincts.
}
\label{tab:4_runtimes}
\resizebox{13cm}{!} {
\begin{tabular}{l l l l l l l l l}
\toprule
\multirow{2}{*}{County} & \multicolumn{2}{c}{VR} & \multicolumn{2}{c}{Alpha} & \multicolumn{2}{c}{Adjacency} & \multicolumn{2}{c}{Level-set} \\
\cmidrule(lr){2-3} \cmidrule(lr){4-5} \cmidrule(lr){6-7} \cmidrule(lr){8-9} 
& Complex & PH & Complex & PH & Complex & PH & Complex & PH \\
\midrule
El Dorado (T) & 180.426 s & 0.783 s & -- & -- & \textbf{0.090 s}& \textbf{0.008 s} & 2.580 s & 0.011 s \\
Imperial (C) & 0.739 s & 0.154 s & -- & -- & \textbf{0.0137 s} & 0.009 s & 9.29 s & \textbf{0.007 s} \\
Los Angeles (C) & -- & -- & 15.479 s & 0.065 s & 39.264 s & 0.069 s & \textbf{8.842 s} & \textbf{0.045 s} \\
Merced (C) & 488.823 s & 0.669 s & -- & -- & \textbf{0.0217 s} & \textbf{0.009 s} & 6.677 s & 0.025 s \\
Napa (C) & 654.803 s & 0.980 s & -- & -- & \textbf{0.048 s} & \textbf{0.010 s} & 9.161 s & 0.042 s \\
San Bernardino (C) & -- & -- & 1.765 s & 0.032 s & \textbf{0.691 s} & 0.030 s & 4.385 s & \textbf{0.019 s} \\
Tulare (T) & -- & -- & \textbf{0.0515 s} & 0.016 s & 0.129 s & 0.015 s & 5.180 s & \textbf{0.006 s} \\
\bottomrule
\end{tabular}
}
\end{table}

\subsection{Barcodes and Feature Maps}
\label{ss:resultsloops}

In this section, we use examples to illustrate the differences between the results of the various methods. We generate two types of visualizations for our PH results. The first takes the form of barcodes, where we display each feature as a bar whose length corresponds to its persistence. The second is a map visualization, where we mark the locations of the features that we find by computing PH by drawing a cycle that passes through all of the generators of a feature. (We call this a ``feature map''.) These generators are not necessarily unique, and we select our generators by using a standard PH algorithm (specifically, by using the row-reduced boundary matrix) \cite{Zomorodian2005}. Although the non-uniqueness of generators is a potential concern, in our study, any set of generators results in some group of precincts that surround a voting island. We color the cycle according to the political party of the candidate. For example, if we find a blue hole in a sea of red, we draw a red cycle. To help illustrate the various interpretations of persistence, we highlight ``long-persistence'' features in $H_1$. Specifically, if an element $[x] \in H_1$ has persistence interval $[\text{birth}([x]),\text{death}([x])]$, we compute
\begin{equation}
	l = \frac{\text{death}([x])-\text{birth}([x])}{\max_{[y] \in H_1} \left[\text{death}([y]) - \text{birth}([y])\right]}\,.
\end{equation}
If $l \geq 0.75$, we consider $[x]$ to be a long-persistence feature. We color long-persistence features in dark red or dark blue, depending on the political party of the candidate, and we color other features in lighter shades of red or blue. We also color long-persistence features with darker bars in the barcodes. We discuss results for two counties in this section, and we give additional examples in Appendix~\ref{app:ex}. 

For our first example, we compare the barcodes and feature maps that we obtain by computing PH of the alpha, adjacency, and level-set complexes that we generate from red precincts (i.e., those with a majority who voted for Donald Trump) in Tulare County (see Figure~\ref{fig:4_tulare}). Tulare County is relatively small, with only $250$ precincts. The county is predominantly rural, although it has a few small urban areas towards its western side. Politically, Tulare is a strongly Republican county, with only a very small proportion of its precincts who voted blue (i.e., for Hillary Clinton) in the 2016 election. Looking at a voting map of Tulare, we observe several pockets of blue voters that we hope to be able to detect using PH. To detect these blue pockets, we consider the topological structure of simplicial complexes that we construct using only the part of the map with red precincts, and we seek to find holes in these complexes. In Figure~\ref{fig:4_tularebar}, we show the results of the three different constructions.

For the alpha complex, we observe that the dimension-$1$ barcodes indicate that most features do not have long persistences. The loops that surround the blue holes are light red, indicating that they are not long-persistence features. Additionally, the single long-persistence feature corresponds to a loop in the northwest part of the voting map; it connects several precincts whose union is disconnected, and it does not surround any blue areas. It thus exhibits both the scaling and contiguity problems that we discussed in Section~\ref{ss:sccomp}. The spacing of these three precincts is such that the pairwise distances between them are similar, but this spacing is at a distance that is larger than the precincts themselves, causing them to form a loop even though none of them are not adjacent to each other on the map. Because this loop corresponds to the only long-persistence bar in the barcode, it is difficult to use persistence to distinguish fake loops like this one from real loops in the western region of the map. Overall, the alpha complex does detect some pockets (of voting results that are surrounded by different voting results), but it misses a few of them just southeast of the central area; it also detects many features that are not real.

In contrast to our observations with the alpha complex, generator precincts in the adjacency complex mostly form contiguous loops; by virtue of our construction, edges cannot occur between the centroids of precincts that are not adjacent to each other. A few features that are disconnected from their (graph-theoretic) neighbors do still appear on the resulting feature map, largely because the precincts themselves have complicated shapes. For example, some of them are not simply connected and others have multiple connected components. Some work in mathematical gerrymandering has focused on tackling some of these issues by quantifying the idea that electoral districts ought to be ``compact'' \cite{DBLP:journals/corr/abs-1803-02857, 2018arXiv180805860D}. However, for the most part, the generator precincts surround blue and light-red holes in the voting map. Additionally, there are fewer bars in the dimension-$1$ barcode in the adjacency complex than in the alpha complex, and more of the bars in the adjacency complex correspond to long-persistence features. The longest bar corresponds to the large hole in the middle that includes both blue and light-red precincts. Although these light-red precincts do eventually joint the filtered simplicial complex, the blue precincts in the middle ensure that this hole never closes. Keeping in mind that the generators of a feature are not necessarily unique, the particular algorithm that we use to compute PH selects the group of darker red precincts that surround that area. We also observe several small light-red holes (which correspond to early-birth bars) and several blue holes (which correspond predominantly to the bars in the barcode that are born late). The adjacency complex is able to locate most of the blue areas of the voting map---the exceptions are a few areas near the edges (and there is no hope of detecting several of these as holes, because they lie on the county's borders and thus cannot be surrounded)---and it has little noise. All of the long-persistence feature are true features, and we can therefore do a better job of distinguishing signal from noise for Tulare County with the adjacency complex than with the alpha complex.

Finally, we examine the level-set barcode and feature map. Observe that the dimension-$1$ barcode has several features---some with long persistence and others without it---that start at time $0$; there is one feature that starts at a much later step of the filtration. These bars correspond to several of the holes in the western area of the voting map. We detect only six of these holes, as some of them occur on size scales that are too small for us to capture in our level-set complex because of our choice of grid resolution during construction. We also observe that the persistence of a bar is correlated positively with the size of its associated hole. The single long-persistence feature corresponds to the largest blue hole. Overall, the level-set complex captures most of the blue areas on the map and avoids most of the noise, although it does fail to detect some of the smaller regions.

\begin{figure}[!htbp]
\centering
\includegraphics[width=.80\textwidth]{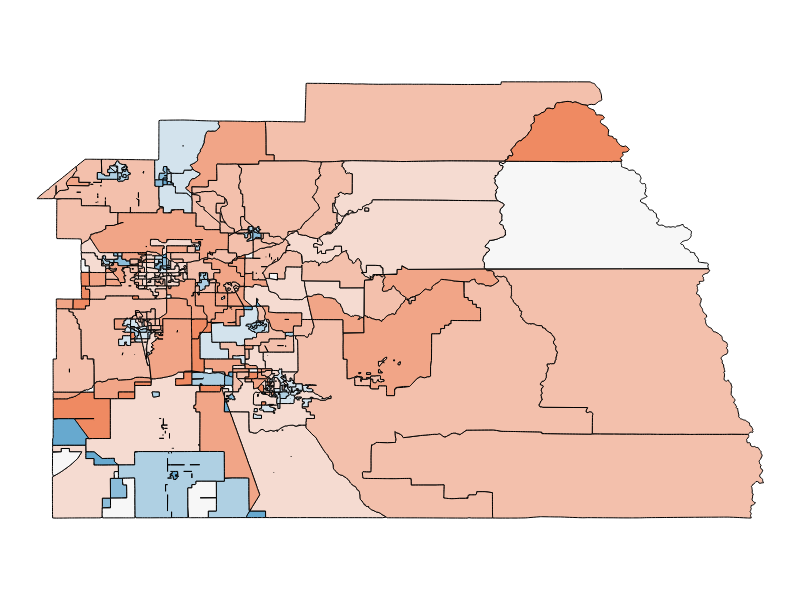}
\vspace{-1.25cm}
\caption{Tulare County, which we color based on the voting for president in the 2016 election. Red precincts have a majority who voted for Trump, and blue precincts have a majority who voted for Clinton. Darker colors indicate stronger majorities.
}
\label{fig:4_tulare}
\end{figure}

\begin{figure}[!htbp]
\centering
\begin{subfigure}[t]{\textwidth}
	\centering
	\includegraphics[width=.45\textwidth]{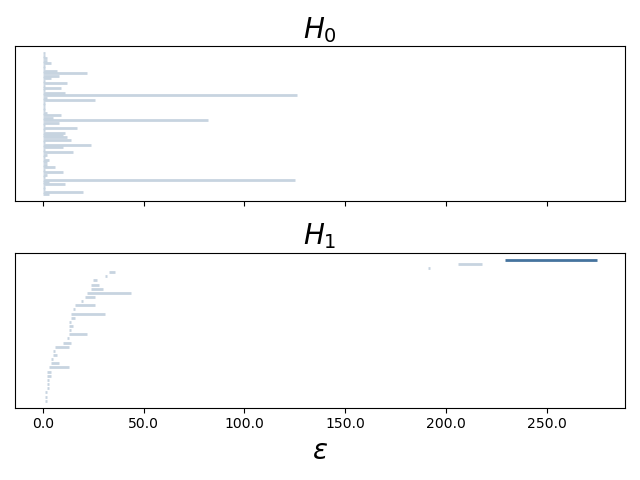}
	\includegraphics[width=.45\textwidth]{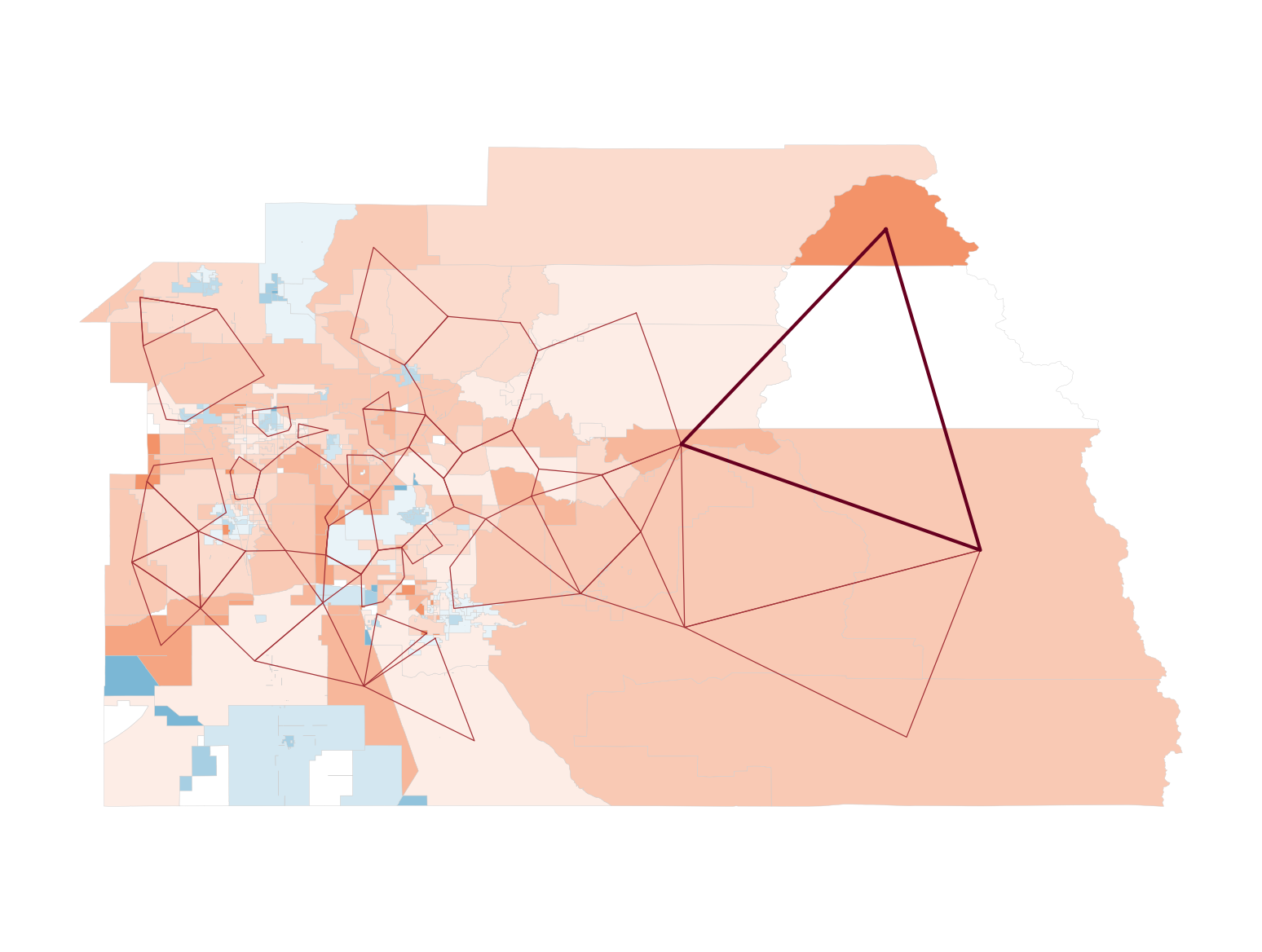}
	\vspace{-.5cm}
	\caption{\label{subfig:4_tulare_a}Alpha complex}
\end{subfigure}
\begin{subfigure}[t]{\textwidth}
	\centering
	\includegraphics[width=.45\textwidth]{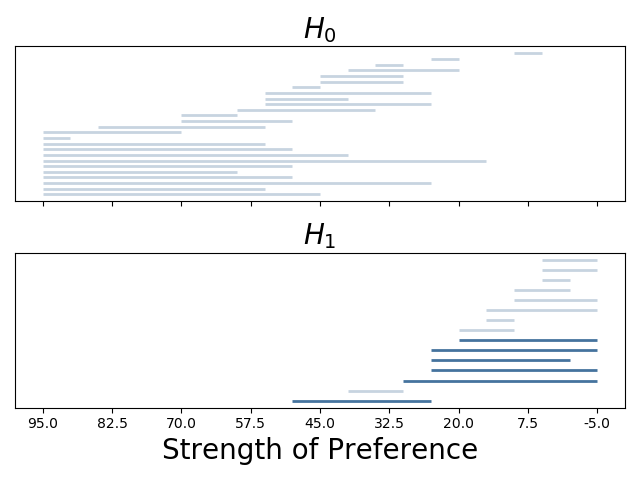}
	\includegraphics[width=.45\textwidth]{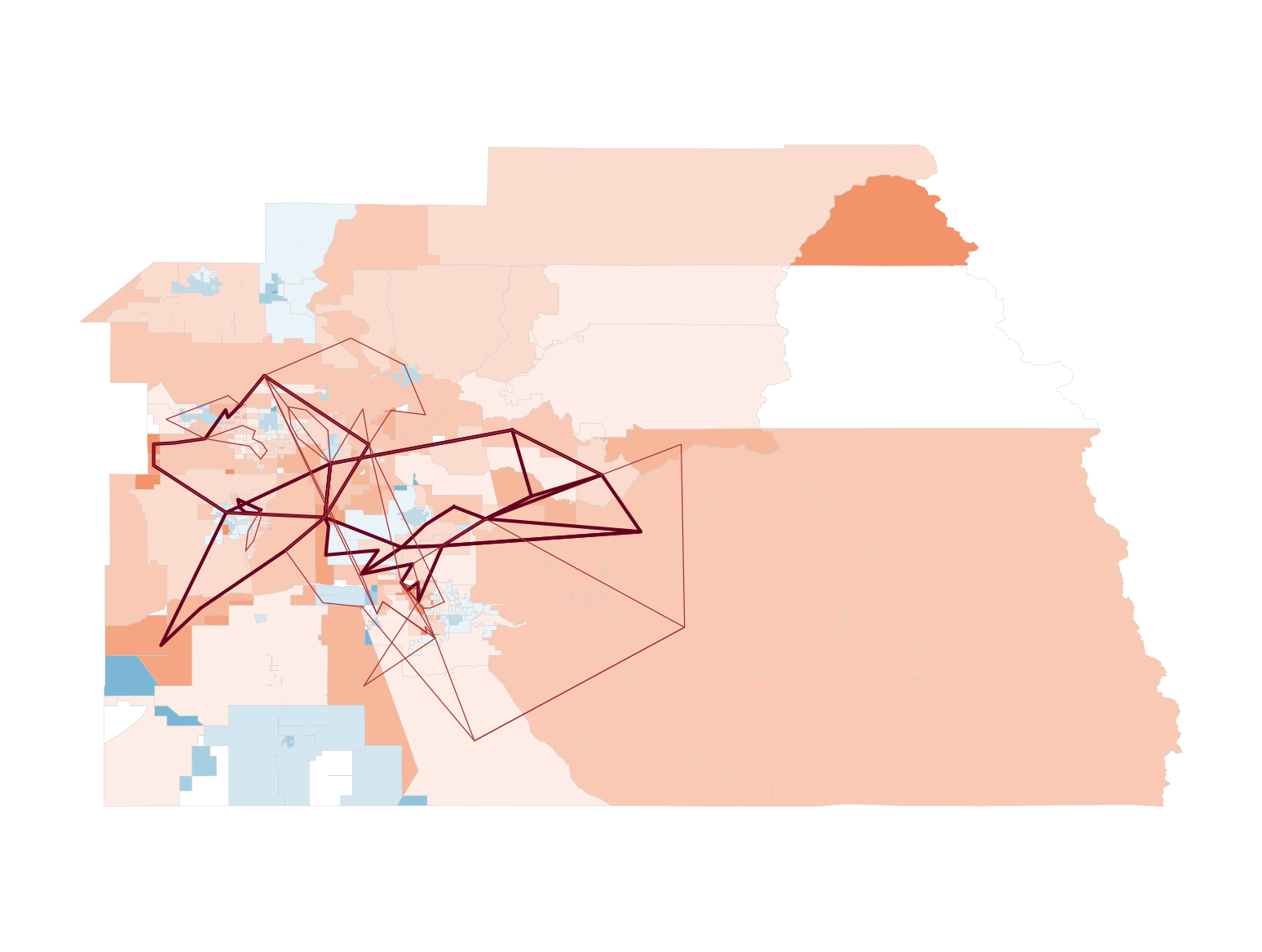}
	\vspace{-.25cm}
	\caption{\label{subfig:4_tulare_b}Adjacency complex}
\end{subfigure}
\begin{subfigure}[t]{\textwidth}
	\centering
	\includegraphics[width=.45\textwidth]{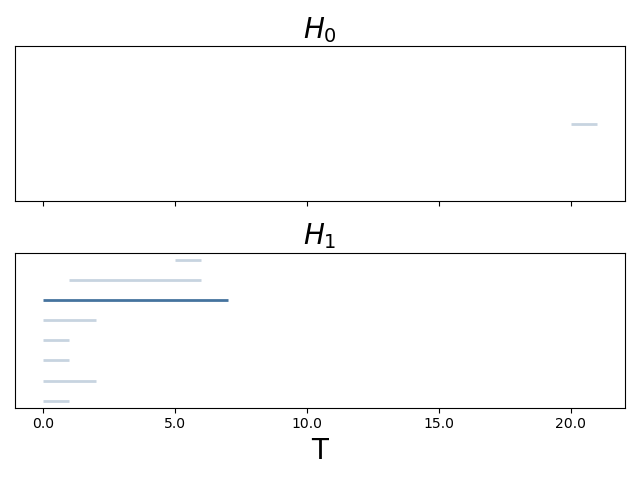}
	\includegraphics[width=.45\textwidth]{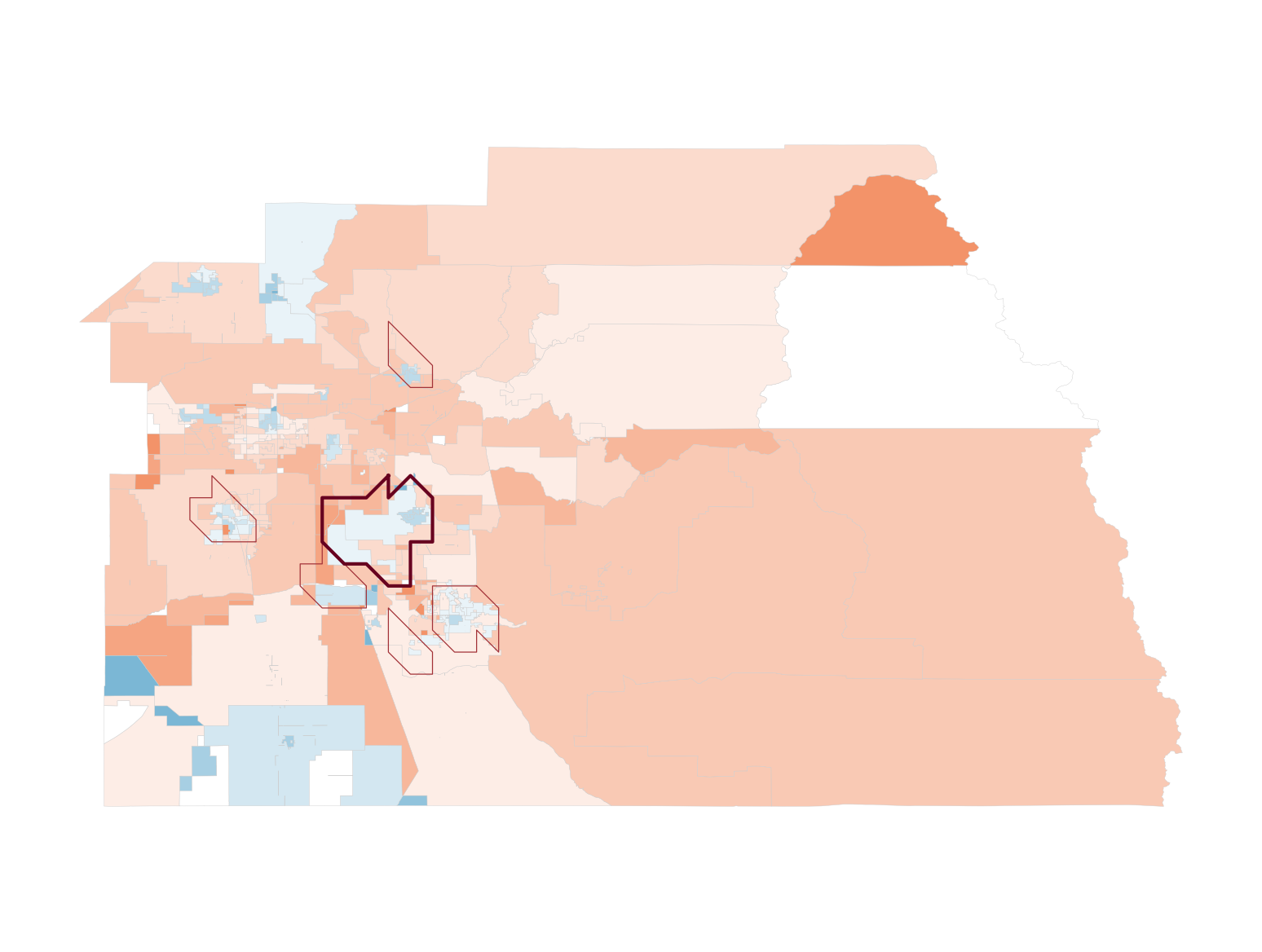} 
	\vspace{-.5cm}
	\caption{\label{subfig:4_tulare_c}Level-set complex}
\end{subfigure}
\vspace{-.5cm}
\caption{Barcodes and generated loops for red precincts in Tulare County. We mark long-persistence features using darker loops with thicker line widths.
}
\label{fig:4_tularebar} 
\end{figure}

For our second example, we consider Imperial County's blue precincts, which we show in the map in Figure~\ref{fig:4_imp}. For a visualization of the various simplicial complexes that we built with Imperial County's red precincts, see Section~\ref{sec:methods}. In contrast to Tulare County, it is not immediately evident for Imperial County where there may be holes. There do seem to be a few very small red precincts that are surrounded by blue precincts, so we hope to be able to capture some of those. Overall, however, we expect to observe relatively few features.

Examining the results from the various constructions, we observe that the VR complex picks up some noise, and only one of the features appears to surround a hole. Instead, it finds several areas where the blue precincts are tightly clustered, but they do not seem to surround any red precincts. Furthermore, all of the features have similar persistences, and they are all categorized long-persistence features. Unfortunately, because so many of the precincts in Imperial County are small, it is unsurprising that all of the features have similar persistences, so it is difficult to distinguish signal from noise. Moreover, as we will see, our findings from the adjacency complex and level-set complex imply that the VR complex is not picking up any real holes.

The adjacency complex picks up one long-persistence feature and two other features. On inspection, these appear to be small white or light-blue holes that are surrounded by darker blue districts. All three of the holes appear to be around either white precincts or red precincts, and the single long-persistence feature is composed of relatively dark-blue precincts. The long-persistence feature also seems to be the only feature that corresponds to a feature from the VR construction.

In contrast to the adjacency and VR complexes, which include very few features, the level-set complex picks up a large number of dimension-$1$ features, but none of them start at time $0$. This occurs because, as the level set evolves, the separate connected components eventually combine, creating a larger number of holes than the ones that actually exist in the original voting map. This illustrates one of the problems with the level-set complex: as time passes, the simplicial complex tends toward becoming progressively more connected, which can create some false features when the simplicial complex starts with many connected components. However, if one considers only those features that exist at time $0$, one can distinguish between genuine and false features. Most of the counties have relatively homogeneous voting patterns, with small pockets of dissimilarity, so few of the California counties exhibit this behavior in practice. Additionally, including only features that begin at time $0$ results in reasonable feature maps.

\begin{figure}[!htbp]
\centering
\includegraphics[width=\textwidth]{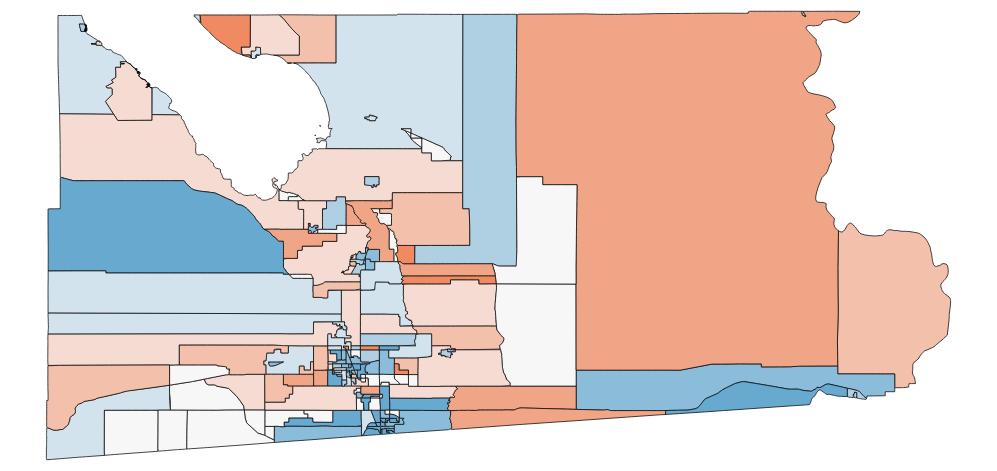}
\caption{Imperial County, which we color based on presidential voting. Red precincts have a majority who voted for Trump, and blue precincts have a majority who voted for Clinton. Darker colors indicate stronger majorities.
}
\label{fig:4_imp}
\end{figure}

\begin{figure}[!htbp]
\centering
\begin{subfigure}[t]{\textwidth}
	\centering
	\includegraphics[width=.45\textwidth]{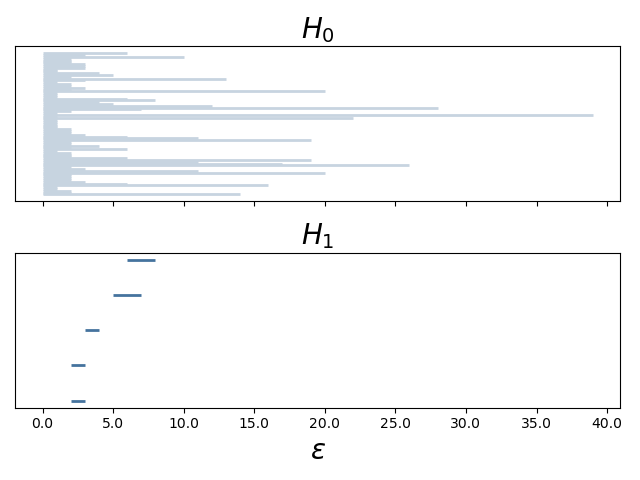}
	\includegraphics[width=.45\textwidth]{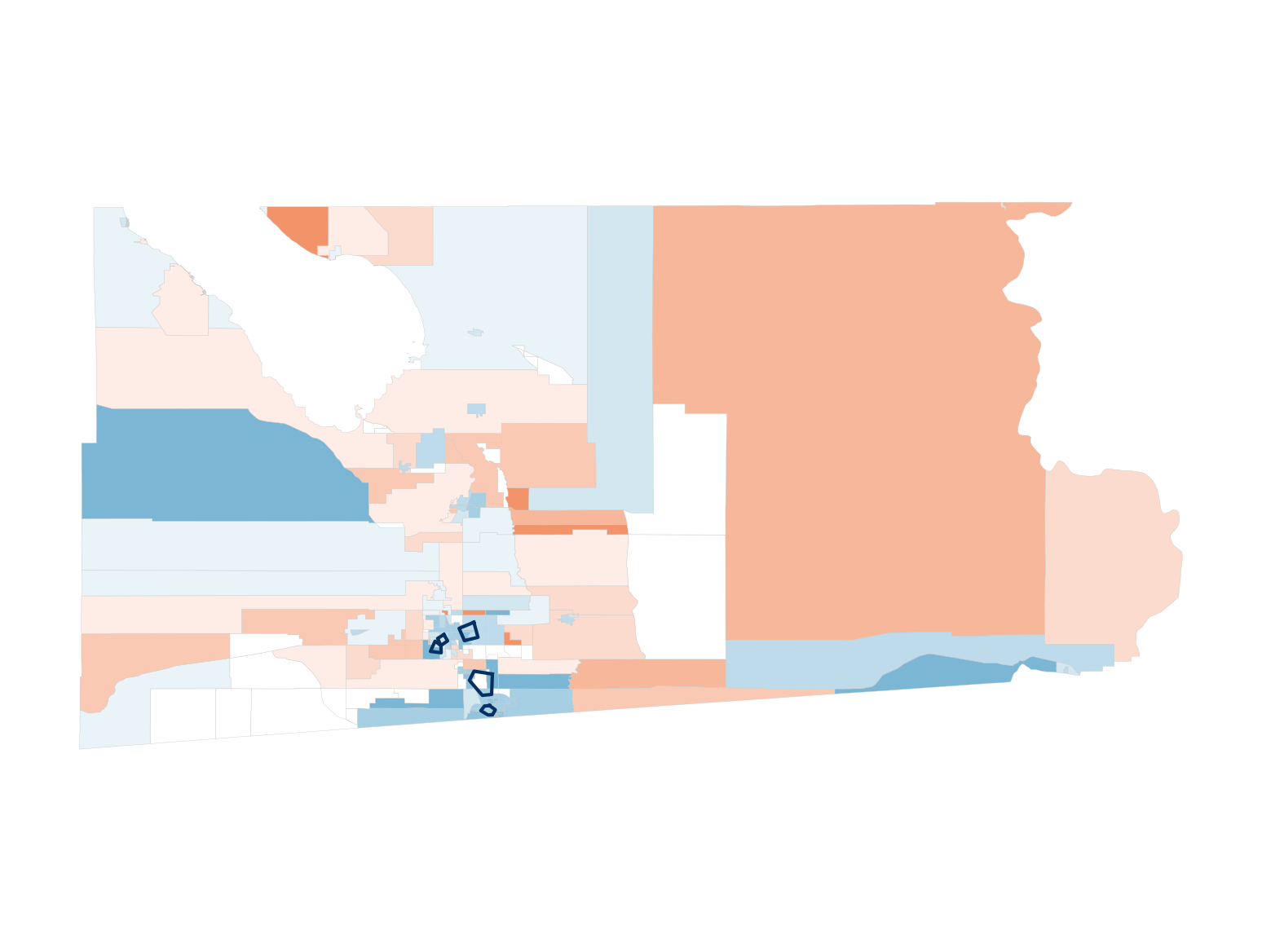}
	\vspace{-.5cm}
	\caption{\label{subfig:4_imp_a}VR complex}
\end{subfigure}
\begin{subfigure}[t]{\textwidth}
	\centering
	\includegraphics[width=.45\textwidth]{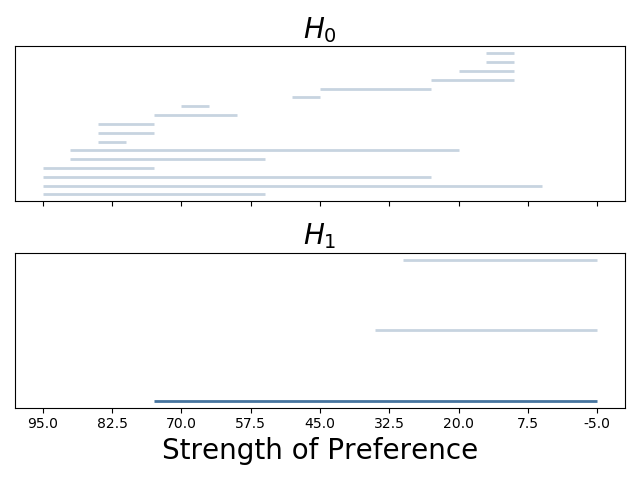}
	\includegraphics[width=.45\textwidth]{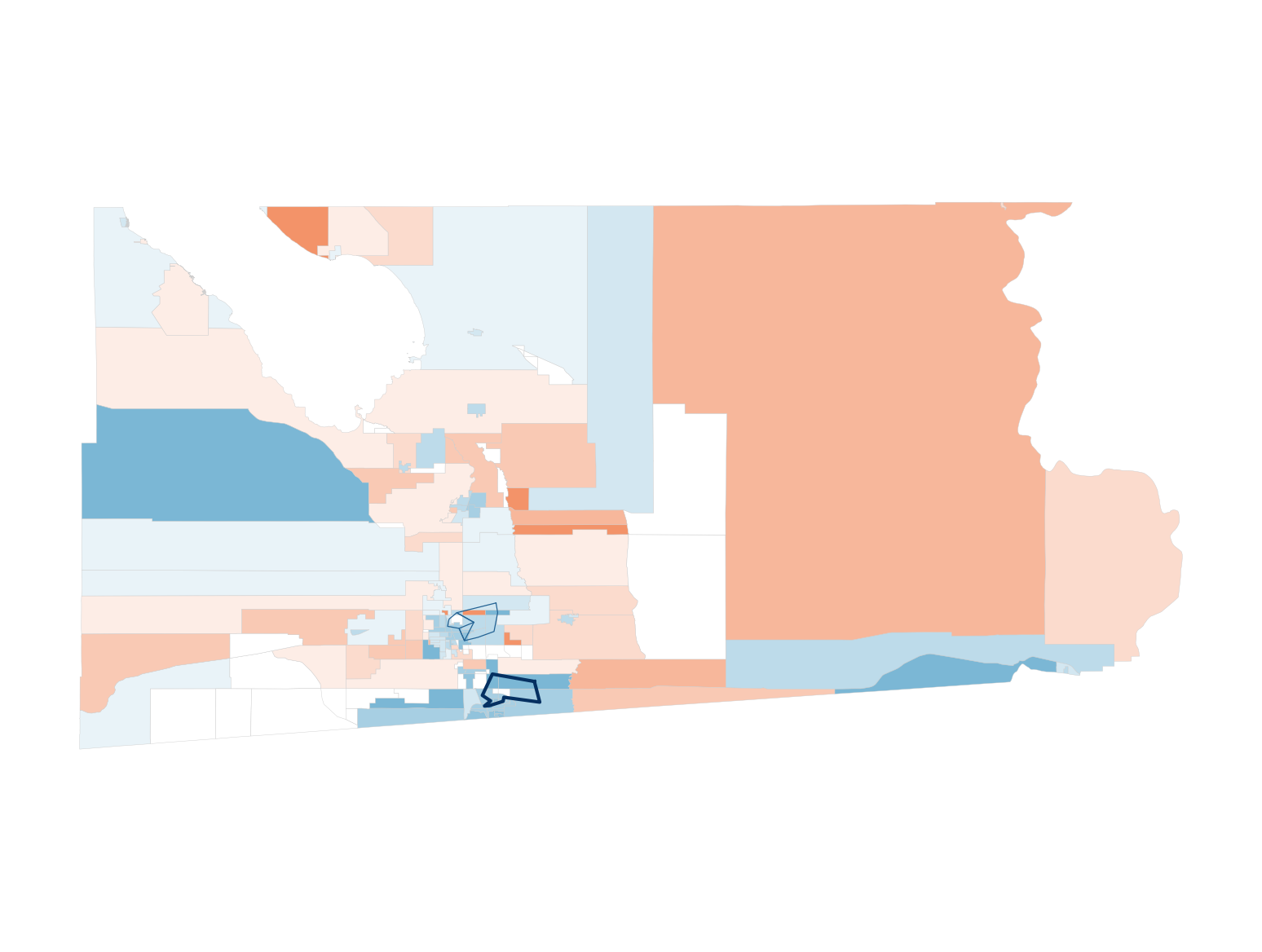}
	\vspace{-.25cm}
	\caption{\label{subfig:4_imp_b}Adjacency complex}
\end{subfigure}
\begin{subfigure}[t]{\textwidth}
	\centering
	\includegraphics[width=.45\textwidth]{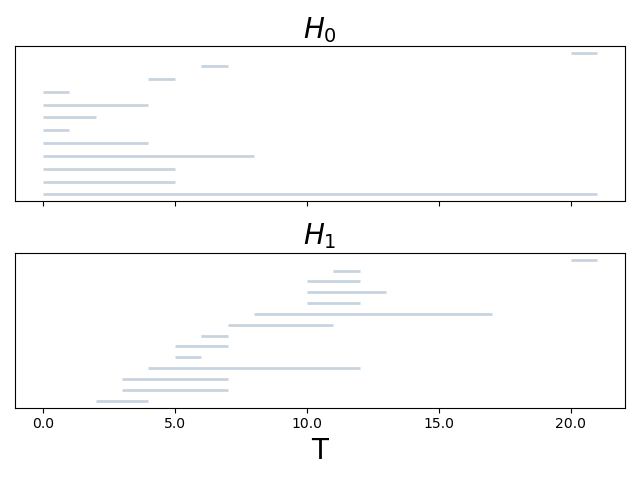}
	\includegraphics[width=.45\textwidth]{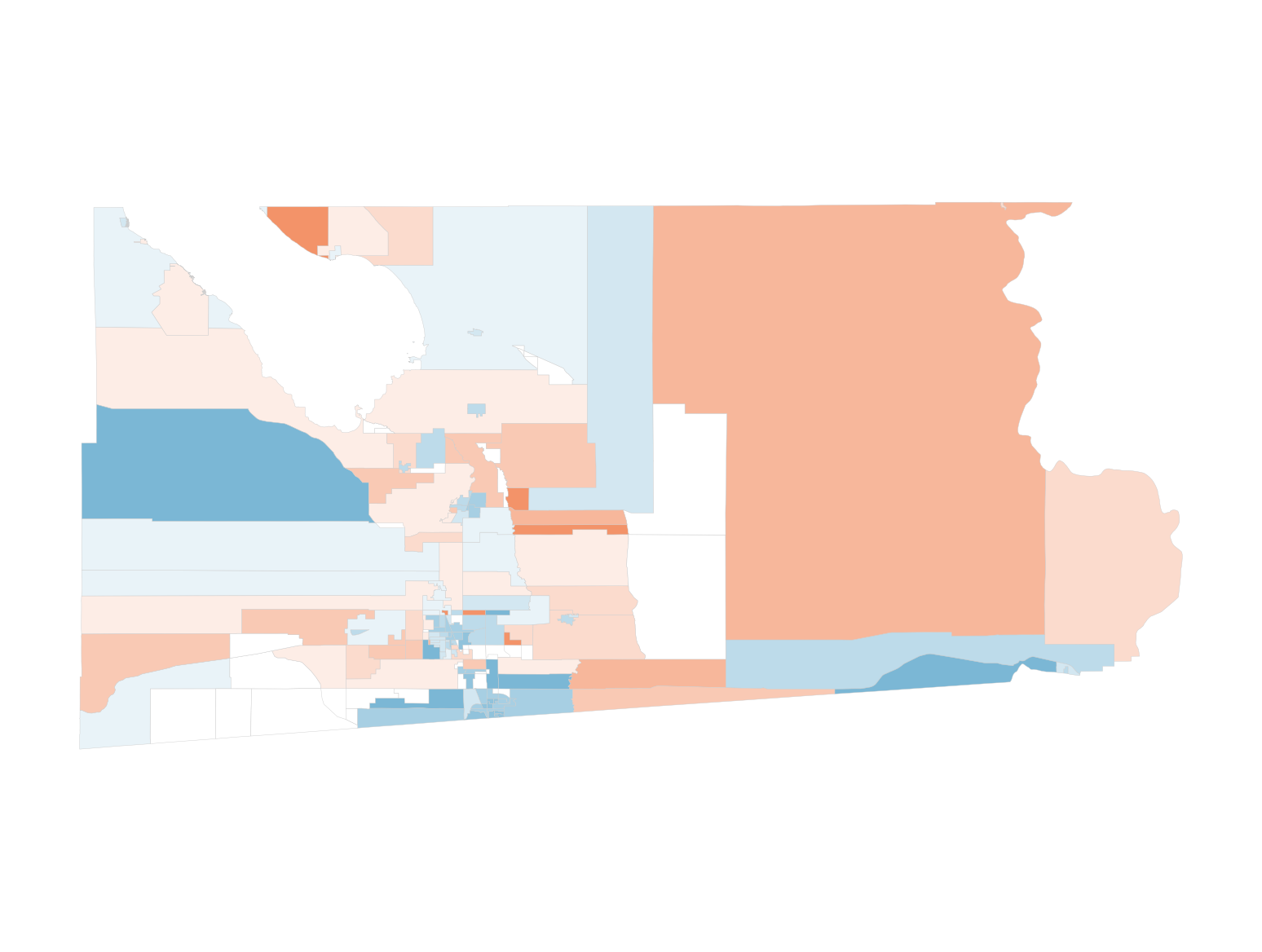} 
	\vspace{-.5cm}
	\caption{\label{subfig:4_imp_c}Level-set complex}
\end{subfigure}
\caption{Barcodes and generated loops for blue precincts in Imperial County. The VR complex results in several false ``features''; the adjacency complex detects two white holes and one red hole; and the level-set complex is unable to detect any holes, because there do not exist sufficiently large white or red holes.
}
\label{fig:4_impbar} 
\end{figure} 

%%%%%

\subsection{Comparison of Our Results to ``Ground Truth''}\label{ss:gt}

We conclude our analysis with some discussion of the accuracy with which we are able to use long-persistence features to find true features in the \emph{LA Times} voting data. In Table~\ref{tab:5_gt}, we show the proportion of long-persistence features that indicate an actual hole, as determined by the human eye. We highlight the most successful method for each county in bold. We see that our adjacency and level-set approaches outperform the VR and Alpha constructions. This indicates that our methods are less likely than the traditional distance-based approaches to detect noise as significant features in these examples.

\begin{table}[!htbp]
\tiny
\centering
\caption{Proportion of long-persistence features that identify a real feature in our simplicial complexes.}
\label{tab:5_gt}
\resizebox{12cm}{!} {
\begin{tabular}{l l l l l l l l l l}
\toprule
\multirow{2}{*}{County}  & \multicolumn{2}{c}{VR} & \multicolumn{2}{c}{Alpha} & \multicolumn{2}{c}{Adjacency} & \multicolumn{2}{c}{Level-set} \\
\cmidrule(lr){2-3}
\cmidrule(lr){4-5}
\cmidrule(lr){6-7}
\cmidrule(lr){8-9}

& C & T & C & T & C & T & C & T \\
\midrule

Alameda & -- & 0.00 & {\bf 1.00} & -- & {\bf 1.00} & -- & {\bf 1.00} & -- \\
Alpine & -- & --  & -- & -- & -- & -- & -- & -- \\
Amador  & -- & {\bf 1.00} & -- & -- & -- & -- & -- & -- \\
Calaveras  & -- & {\bf 1.00} & -- & -- & -- & -- & -- & {\bf 1.00} \\
Colusa  & -- & {\bf 1.00} & -- & -- & -- & -- & -- & {\bf 1.00} \\
Contra Costa  & -- & 0.00 & 0.00 & -- & {\bf 1.00} & -- & {\bf 1.00} & {\bf 1.00} \\
Del Norte  & -- & 0.00 & -- & -- & -- & {\bf 1.00} & -- & 0.00 \\
El Dorado  & 0.00 & {\bf 1.00} & -- & -- & {\bf 1.00} & {\bf 1.00} & -- & {\bf 1.00} \\
Fresno  & -- & -- & 0.00 & 0.00 & 0.66 & 0.00 & -- & {\bf 1.00} \\
Glenn  & -- & 0.00 & -- & -- & -- & 0.00 & -- & {\bf 1.00} \\
Humboldt  & 0.00 & 0.00 & -- & -- & 0.50 & -- & {\bf 1.00} & {\bf 1.00} \\
Imperial  & 0.20 & {\bf 1.00} & -- & -- & {\bf 1.00} & {\bf 1.00} & -- & {\bf 1.00} \\
Inyo  & -- & 0.00 & -- & -- & -- & {\bf 1.00} & -- & -- \\
Kern  & -- & -- & 0.00 & {\bf 1.00} & {\bf 1.00} & {\bf 1.00} & -- & {\bf 1.00} \\
Kings  & 0.00 & 0.00 & -- & -- & {\bf 1.00} & 0.67 & -- & 0.87 \\
Lake  & {\bf 1.00} & 0.00 & -- & -- & -- & -- & {\bf 1.00} & -- \\
Lassen  & -- & {\bf 1.00} & -- & -- & -- & -- & {\bf 1.00} & {\bf 1.00} \\
Los Angeles  & -- & -- & 0.00 & 0.00 & -- & -- & {\bf 1.00} & -- \\
Madera  & {\bf 1.00}& {\bf 1.00} & -- & -- & {\bf 1.00} & {\bf 1.00} & -- & {\bf 1.00} \\
Marin & -- & -- & {\bf 1.00} & -- & {\bf 1.00} & -- & {\bf 1.00} & -- \\
Mariposa  & -- & {\bf 1.00} & -- & -- & -- & -- & -- & -- \\
Mendocino  & -- & 0.00 & {\bf 1.00} & -- & {\bf 1.00} & -- & {\bf 1.00} & -- \\
Merced  & 0.11 & {\bf 1.00} & -- & -- & 0.5 & {\bf 1.00} & -- & {\bf 1.00} \\
Modoc & -- & 0.00  & -- & -- & --  & -- & -- & --\\
Mono  & 0.00 & -- & -- & -- & -- & -- & -- & -- \\
Monterey  & -- & 0.00 & 0.00 & -- & {\bf 1.00} & 0.00 & {\bf 1.00} & {\bf 1.00} \\
Napa  & 0.25 & 0.00 & -- & -- & {\bf 1.00} & -- & .75 & -- \\
Nevada  & 0.00 & {\bf 1.00} & -- & -- & {\bf 1.00} & {\bf 1.00} & {\bf 1.00} & {\bf 1.00}\\
Orange  & -- & -- & 0.00 & 0.00 & 0.00 & 0.50 & {\bf 1.00} &{\bf 1.00}\\
Placer  & 0.50 & -- & -- & 0.00 & -- & {\bf 1.00} & {\bf 1.00} & {\bf 1.00} \\
Plumas  & -- & {\bf 1.00} & -- & -- & -- & {\bf 1.00} & -- & {\bf 1.00} \\
Riverside  & -- & -- & 0.00 & .33 & {\bf 1.00} & {\bf 1.00} &{\bf 1.00} & {\bf 1.00}\\
Sacramento & -- & -- & 0.00 & 0.00 & 0.00 & {\bf 1.00} & {\bf 1.00} & {\bf 1.00} \\
San Benito & {\bf 1.00} & 0.00 & -- & -- & {\bf 1.00} & -- & -- & {\bf 1.00} \\
San Bernardino  & -- & -- & 0.00 & 0.00 & -- & 0.75 & -- & {\bf 1.00} \\
San Diego  & -- & -- & 0.00 & {\bf 1.00} & {\bf 1.00} & {\bf 1.00} & {\bf 1.00} & {\bf 1.00} \\
San Francisco  & -- & -- & 0.00 & -- & {\bf 1.00} & -- & {\bf 1.00} & -- \\
San Joaquin  & -- & -- & 0.00 & 0.00 & 0.75 & {\bf 1.00} & {\bf 1.00} & {\bf 1.00} \\
San Luis Obispo  & 0.00 & 0.14 & -- & -- & {\bf 1.00} & {\bf 1.00} & -- & {\bf 1.00} \\
San Mateo  & -- & -- & {\bf 1.00} & -- & {\bf 1.00} & -- & {\bf 1.00} & -- \\
Santa Barbara  & -- & {\bf 1.00} & 0.00 & -- & .67 & {\bf 1.00} & -- & {\bf 1.00} \\
Santa Cruz  & -- & -- & {\bf 1.00} & -- & 0.00 & -- & {\bf 1.00} & -- \\
Shasta  & -- & 0.00 & -- & -- & -- & {\bf 1.00} & -- &-- \\
Sierra  & -- & -- & -- & -- & -- & -- & -- & -- \\
Solano  & 0.00 & 0.00 & -- & -- & {\bf 1.00} & {\bf 1.00} & {\bf 1.00} & {\bf 1.00} \\
Sonoma  & -- & 0.00 & 0.00 & -- & {\bf 1.00} & -- & {\bf 1.00} & -- \\
Stanislaus  & 0.00 & 0.00 & -- & -- & {\bf 1.00} & {\bf 1.00} & -- & {\bf 1.00} \\
Sutter& -- & 0.00 & -- & -- & -- & {\bf 1.00} & -- & {\bf 1.00}\\
Tehama &-- & 0.00 & -- & -- & -- & {\bf 1.00} & -- & -- \\
Trinity  & -- & 0.00 & -- & -- & -- & 0.00 & {\bf 1.00} & {\bf 1.00} \\
Tulare  & 0.00 & -- & -- & 0.00 & -- & {\bf 1.00} & -- & {\bf 1.00} \\
Tuolomne  & -- & 0.00 & -- & 0.00 & -- & {\bf 1.00} & -- & {\bf 1.00} \\
Yolo  & 0.00 & -- & -- & -- & {\bf 1.00} & {\bf 1.00} & 0.00 & 0.00 \\
Yuba & -- & 0.00 & -- & -- & -- & {\bf 1.00} & -- & {\bf 1.00} \\
\bottomrule
\end{tabular}
}
\end{table}

%%%%%

\section{Conclusions}
\label{sec:conclusion}

Analyzing persistent homology in geospatial data can often lead to results that are difficult to interpret because of the heterogeneity of distance scales in such data. A particularly difficult aspect is that barcodes of a similar length may represent either signal or noise, in stark contrast to the conventional wisdom that the features that persist the longest also carry the most meaningful information about a data set. The difficulty in identifying interesting features from a barcode can make PH a challenging tool to apply effectively, even in applications in which topological holes seem like something that is appropriate to compute to gain insights into a problem. Therefore, it is extremely important to further explore the issue signal versus noise in PH, especially for multiscale problems. In this paper, we introduced two new methods for constructing a filtered simplicial complex that approximates a geographic map, and we discussed the effects that different types of complexes have on the resulting PH. Our approaches attempt to address the difficulties of applying topological data analysis (TDA) to data that is not well-represented by traditional point clouds. Our adjacency complex allowed us to incorporate data about relationships other than distance between points, while preserving the embedding of the geographic map in space and without having to make specific choices of distance transformations for different counties. Our level-set complex allowed us to compute, in a relatively inexpensive way, complexes that are very similar in intuition to the traditional VR constructions without having to start from a point cloud.  

Both the adjacency and level-set complexes do a better job than traditional distance-based complexes of encoding information about the contiguity of the maps, thereby making it possible to interpret differences in the distance scale of features. An adjacency complex does this by ignoring distance entirely in its construction. For a level-set complex, the persistence of the features that we detect encodes the distance scaling of those features, but with fewer concerns than in VR or alpha complexes about noise due to precinct sizes. Consequently, the barcodes for the adjacency and level-set complexes are more interpretable than those from traditional PH constructions for our geospatial data, allowing us to better understand the topology of voting patterns in counties from the barcodes alone. In future work, it is worth considering adjustments to our constructions that may be helpful for better detecting voting islands. For example, one may wish to apply a scaling based on voting preference (as in our adjacency construction) to a geographical map instead of to precinct vertices to obtain a sub-level-set filtration. Such an approach may help leverage the voting-strength interpretation of the adjacency method while also enjoying the easily interpretable visual contiguity of the level-set simplicial construction.

Although we tailored our methods to yield improvements for the particular problem of detecting voting patterns from {\sc shapefile} data, one can use an adjacency construction on data sets with a network structure, and the level-set construction is appropriate for any type of 2D manifold data (and one can extend it to higher dimensions with some programming adjustments). More generally, given the ubiquity of 2D spatial data, the insights that we highlighted with our voting application are relevant for a broad range of problems, including transportation networks, spatial demography, granular materials, many structures in biology, and others.

%%%%

\appendix

\section{Simplicial Homology} 
\label{app:simplicial}
In this appendix, we discuss the formalism of simplicial homology, which we discussed at an intuitive level in the main text. There are many different homology theories in algebraic topology. We give context for our particular choice of simplicial homology, and we explain some of the differences between simplicial homology and other common homology theories. For more information, see \cite{hatcher2002algebraic}. 

We begin by defining some of the basic building blocks of simplicial homology.

\begin{definition} An \textbf{$n$-simplex} is an $n$-dimensional polytope that is the convex hull of its $n+1$ vertices.
\end{definition}

\begin{definition} An \textbf{orientation} of an $n$-simplex is an ordering of the vertices, written as $(v_0, \ldots, v_k)$, with the rule that two orderings define the same orientation if and only if they differ by an even permutation.
\end{definition}

{\begin{definition} An  \textbf{$m$-face} is the convex hull of a subset of cardinality $m+1$ of an $n$-simplex, with $m<n$ and the orientation preserved. A \textbf{face} refers to an $m$-face of any dimension $m$.
\end{definition}

\begin{definition} A simplex $A$ is a \textbf{coface} of a simplex $B$ if $B$ is a face of $A$.
\end{definition}}

\begin{definition}A \textbf{simplicial complex} $S$ is a set of simplices that satisfies the following conditions:
\begin{enumerate}
	\item{every face of a simplex from $S$ is also in $S$;}
	\item{the intersection of any two simplices $\sigma_1, \sigma_2 \in S$ is a face of both $\sigma_1$ and $\sigma_2$.}
\end{enumerate}
\end{definition}

Our definition of simplicial complex makes no use of orientation. However, in our discussion of simplicial homology, we will see that orientation of simplices is very important.

\begin{definition} Let $S$ be a simplicial complex. A \textbf{simplicial $k$-chain} is a finite formal sum 
\begin{equation*}
	\sum_{i=1}^N c_i \sigma_i\,,
\end{equation*}
where $\sigma_i$ is an oriented $k$-simplex and each $c_i \in F$ for some field $F$.
\end{definition}

We denote the group of $k$-chains on $S$ by $C_k$. (With a consistent choice of orientation, we can also consider this as the free Abelian group on the basis of $k$-simplices in $S$.)

\begin{definition} Let $\sigma = (v_0,\ldots,v_k)$ be an oriented $k$-simplex. The \textbf{boundary operator}
\begin{equation*}
	\delta_k : C_k \to C_{k-1}
\end{equation*}	
is the homomorphism defined by
\begin{equation*}
	\delta_k(\sigma) = \sum_{i=0}^k (-1)^i (v_0, \ldots, \hat{v_i}, \ldots, v_k)\,,
\end{equation*}	
where $(v_0, \ldots, \hat{v_i}, \ldots, v_k)$ is the oriented $(k-1)$-simplex that we obtain by deleting the $i$th vertex of $\sigma$.

Elements of $Z_k = \mathrm{ker}\, \delta_k$ are called \textbf{cycles}, and elements of $B_k = \mathrm{im}\, \delta_{k+1}$ are called \textbf{boundaries}.
\end{definition}

One can show by direct computation that $\delta^2 = 0$, so the groups $(C_k, \delta_k)$ form a chain complex.

\begin{definition} The \textbf{$k$th homology group} $H_k$ of $S$ over $F$ is the quotient group
\begin{equation*}
	H_k(S;F) = Z_k / B_k \,.
\end{equation*}	
\end{definition}

Note that $H_k(S;F)$ is nontrivial precisely when there are $k$-cycles on $S$ that are not boundaries; this occurs when there are $k$-dimensional holes. For example, a cycle between three points gives a $2$-cycle\footnote{This notion of ``cycle'' is somewhat different from the one in network analysis \cite{newman2018}.}, and it is also a boundary precisely when the triangle with vertices at those three points is in the simplicial complex.

In our application (and in many applications of TDA), we compute homology groups over the field $\mathbb{F}_2$. Crucially, $1 = -1 \in \mathbb{F}_2$, so we do not need to consider the orientation of our simplicial complexes.

The final definition that we introduce is that of a simplicial map.

\begin{definition} Let $S$ and $T$ be simplicial complexes. A \textbf{simplicial map} $f: S\to T$ is a function from the vertex set of $S$ to the vertex set of $T$ that preserves simplices. 
\end{definition}

A simplicial map $f: S \to T$ also induces a homomorphism $f_* : H_k(S) \to H_k(T)$ for each integer $k$. The homomorphism $f_*$ is associated with a chain map from the $k$-chain complex of $S$ to the $k$-chain complex of $T$. This chain map is
\begin{equation*}
	(v_0, \ldots, v_k) \mapsto (f(v_0), \ldots, f(v_k))\,,
\end{equation*}	
where $(f(v_0),\ldots,f(v_k)) = 0$ if any of $f(v_0),\ldots, f(v_k)$ are not distinct.

This construction gives a functor from simplicial complexes to Abelian groups; this is essential to the theory of PH that we discussed in Section~\ref{ss:bgph}.

%%%%%

\section{Algorithms and Implementations}
\label{app:alg} 
In this appendix, we discuss the algorithms that we developed to construct our simplicial complexes. All implementations that we discuss in this section 
are available at \url{https://github.com/mhcfeng/precinct}. For the computation of VR and alpha complexes, we used built-in functionality of the software package {\sc Gudhi} \cite{gudhi:FilteredComplexes}. For the adjacency and level-set constructions, we implement (in {\sc Python}) the incremental VR algorithm that is described in \cite{Zomorodian_2010}. This algorithm adds one vertex at a time to a simplicial complex, and it then checks all possible cofaces of that vertex; it adds them if all other vertices of a coface are already part of the simplicial complex. To use this algorithm, we need to do some preprocessing, which we discuss in the next two subsections.

%%%%

\subsection{Adjacency Complex}
\label{app:algadj}

The incremental VR algorithm that we use requires the following items as input: a list of vertices; a list of neighbors for each vertex; and some method of ordering the vertices to compare whether or not a neighbor is a ``lower neighbor'' (i.e., a neighbor that appears prior to the vertex in the ordering). Specifically, the ordering of the vertices must respect the entry times of those vertices. To determine the neighboring precincts for each precinct, we wrote code in QGIS that checks for queen adjacency. (Recall from the main text that two precincts are queen adjacent if they touch each other at any point, including corners.

We then sort precincts by strength of preference for a particular candidate, as the precincts with the strongest preferences enter the filtered simplicial complex first. Once we set this ordering, we compare a precinct to its neighbors to determine whether its neighbors are already in the simplicial complex. It is then straightforward to apply the incremental VR algorithm.

%%%%%

\subsection{Level-Set Complex}
\label{app:algls}

Constructing a level-set complex requires several steps. First, we rasterize our {\sc shapefiles} to obtain geographical maps in image format of all precincts in a county that voted for the same candidate. We denote this image data by $X$, and we constrain these images to have dimension no more than $250 \times 250$. We then define a function $\phi(X,0)$, where $\phi(x,0)$ gives the distance from a point $x \in \mathbb{R}^2$ to the boundary, such that the boundary is the $0$-level set of $\phi(X,0)$. We then implement a level-set method with motion according to normal forces \cite{osher2003} to generate the evolved geographical map $\phi(X,T)$ at each time $T$. To convert $\phi$ to a simplicial complex $S$, we implement Algorithm~\ref{alg:scfromphi}, which takes the following items as input: $\phi(X,T)$; a list $V$ of vertices that are already in the simplicial complex $S$; a list $t$ of entry times for all vertices that are already in $S$; and the current time $T$.

\begin{algorithm}
\caption{Generate ordered vertices from $\phi$.}
\label{alg:scfromphi}
\begin{algorithmic}
\STATE Given $\phi$, $V$, $t$, $T$
\STATE $V' = \{v: v \notin V\,; \,\phi(v,T)<0\,; \, \text{row}(v) = 0 \,\,(\mathrm{mod}\,\, 5)\,, \text{col}(v) = 0 \,\,(\mathrm{mod}\,\, 5)\}$
\FOR{$v \in V'$}
	\STATE $V = V + \{v\}$
	\STATE $t(v) = T$
\ENDFOR
\RETURN $V$, $t$
\end{algorithmic}
\end{algorithm}

As vertices, we use only pixels that are in rows and columns that are multiples of $5$ (see Algorithm~\ref{alg:scfromphi}). This prevents us from having more than $50 \times 50$ potential vertices, which would significantly increase computation time. It also reduces the amount of noise in the barcodes, because holes must be sufficiently large in diameter for us to detect them. Once we have a list of vertices along with their entry times, we generate $1$-simplices using Algorithm~\ref{alg:lsadj}.

\begin{algorithm}
\caption{Generate level-set adjacencies}
\begin{algorithmic}
\label{alg:lsadj}
\STATE Given $V$, height $h$ of image, width $w$ of image
\FOR{$v \in V$}
	\STATE Set $N(v)$ to the set of six possible neighbors of $v$. (These are the four cardinal neighbors, along with the northwest and southeast diagonal neighbors. We limit to six neighbors because this results in a convenient triangulation, and connecting to all eight neighbors would result in non-planarity.)
	\STATE $N(v)= N(v) \bigcap V$
\ENDFOR
\RETURN $N$
\end{algorithmic}
\end{algorithm}

Once we have generated the $1$-simplices, we use the entry times $t$ from Algorithm~\ref{alg:scfromphi} to compare whether or not a neighbor of a given vertex is a lower neighbor in the incremental VR algorithm.

%%%%%

\section{Additional Examples}
\label{app:ex}

In Figures~\ref{fig:C_napabar} and~\ref{fig:C_labar}, we show barcodes and feature maps for Napa County and Los Angeles County, further illustrating some of the problems with barcode interpretability that we discussed in Section~\ref{ss:sccomp}.

\begin{figure}[!htbp]
\centering
\begin{subfigure}[t]{\textwidth}
	\centering
	\includegraphics[width=.45\textwidth]{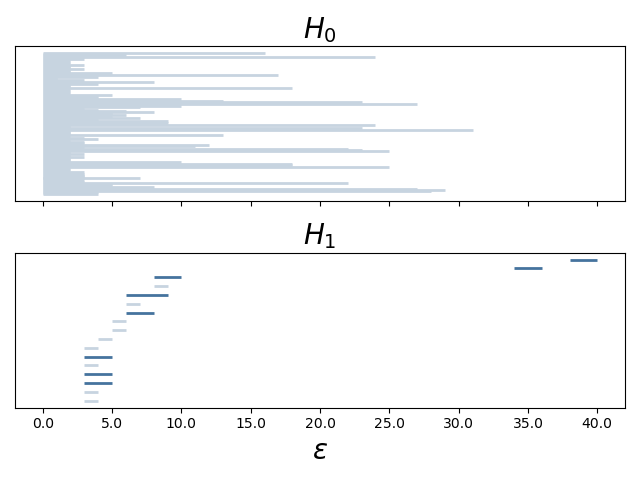}
	\includegraphics[width=.35\textwidth, trim={1cm 0cm 1cm 0cm}, clip]{images/maps/055-napa-ripshillary}
\vspace{-.5 cm}
	\caption{\label{subfig:C_napa_a}VR complex}
\end{subfigure}
\begin{subfigure}[t]{\textwidth}
	\centering
	\includegraphics[width=.45\textwidth]{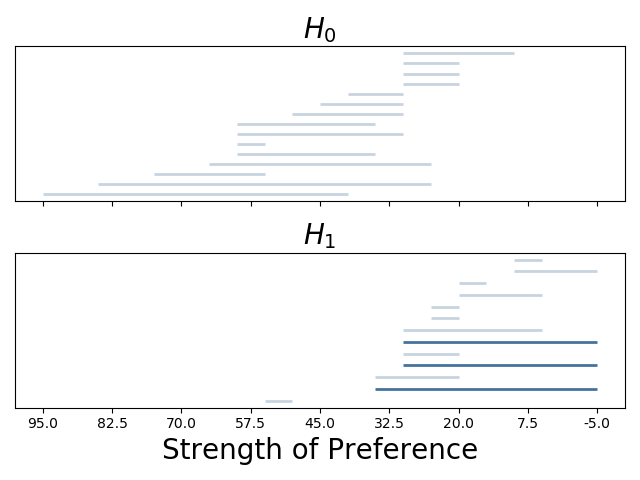}
	\includegraphics[width=.35\textwidth, trim={1cm 0cm 1cm 0cm}, clip]{images/maps/055-napa-adjhillary}
\vspace{-.25 cm}
	\caption{\label{subfig:C_napa_b}Adjacency complex}
\end{subfigure}
\begin{subfigure}[t]{\textwidth}
	\centering
	\includegraphics[width=.45\textwidth]{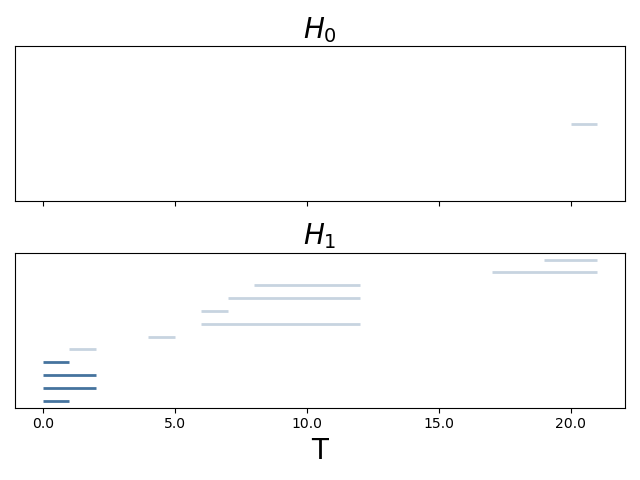}\quad\quad
	\includegraphics[width=.35\textwidth, trim={1cm 0cm 1cm 0cm}, clip]{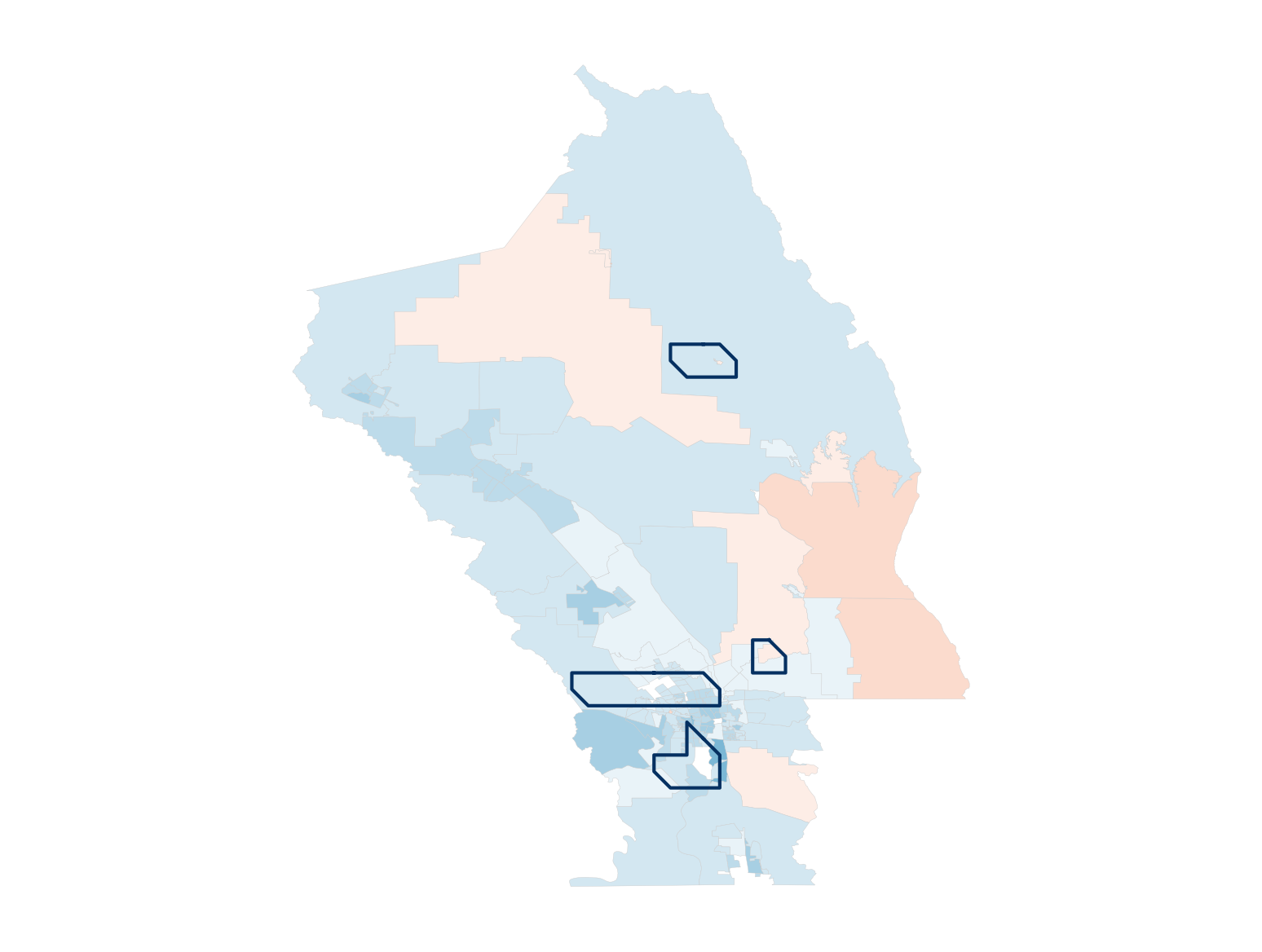} 
\vspace{-.5 cm}
	\caption{\label{subfig:C_napa_c}Level-set complex}
\end{subfigure}
\vspace{-.5 cm}
\caption{\small Barcodes and generated loops for blue precincts in Napa County. There are several long-persistence bars in the $H_1$ barcode of the VR complex. Some of these correspond to real holes in the densely populated areas in the southern region of the county, but others correspond to contiguous blue regions without holes, making it difficult to distinguish signal from noise. By contrast, the $H_1$ barcode for the adjacency complex has three long-persistence features, all of which indicate light-blue or white holes. Similarly, the $H_1$ barcode for the level-set complex has only four features that start at time $0$ and correspond to visible white or red holes. There is a red hole in the eastern part of the region that is detected by the alpha and level-set complexes but not by the adjacency complex. This is due to the shape of the blue precinct, which is wrapped partially around a red precinct such that it covers precisely enough grid points in the level-set complex to register as a hole, despite not actually fully surrounding the red precinct. In practice, this occurs rarely, but it does give an example of a potential problem with the level-set complex.
}
\label{fig:C_napabar} 
\end{figure}

\begin{figure}[!htbp]
\centering
\begin{subfigure}[t]{\textwidth}
	\centering
	\includegraphics[width=.45\textwidth]{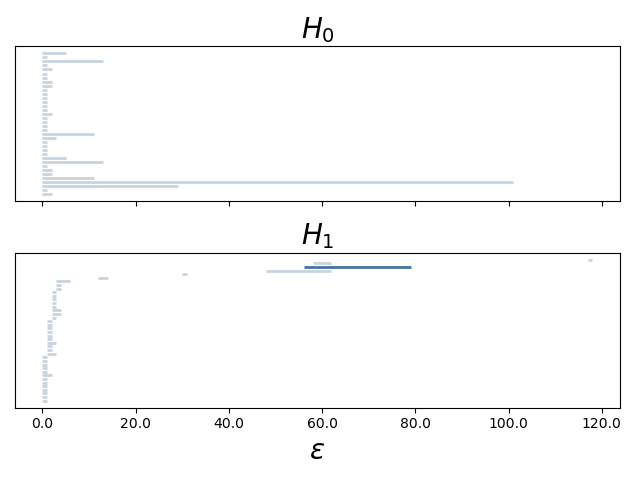}
	\includegraphics[width=.3\textwidth, trim={2cm 0cm 2cm 1cm}, clip]{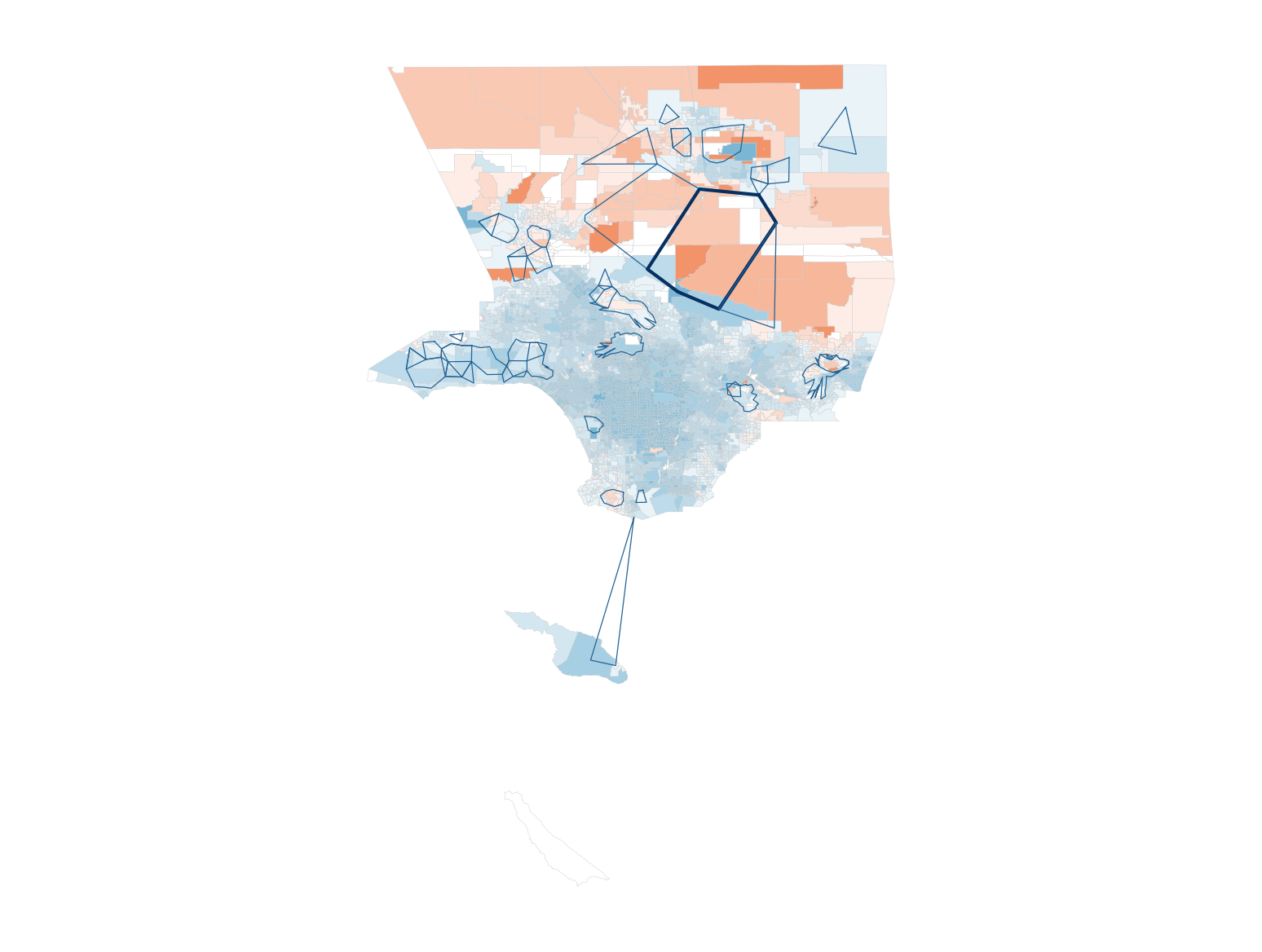}
\vspace{-.5 cm}
	\caption{\label{subfig:C_la_a}Alpha complex}
\end{subfigure}
\begin{subfigure}[t]{\textwidth}
	\centering
	\includegraphics[width=.45\textwidth]{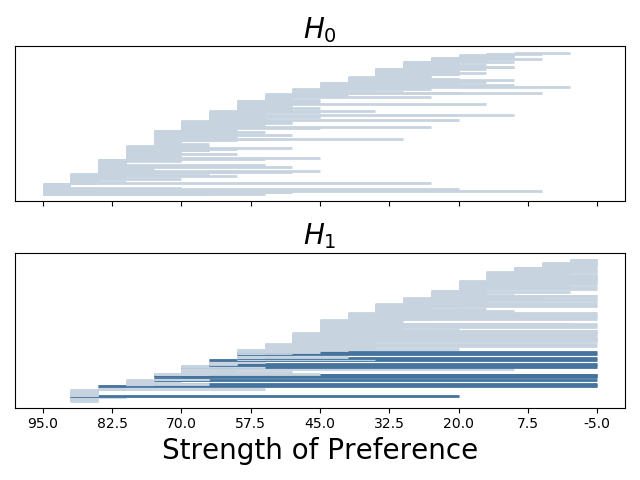}
	\includegraphics[width=.3\textwidth, trim={2cm 0cm 2cm 1cm}, clip]{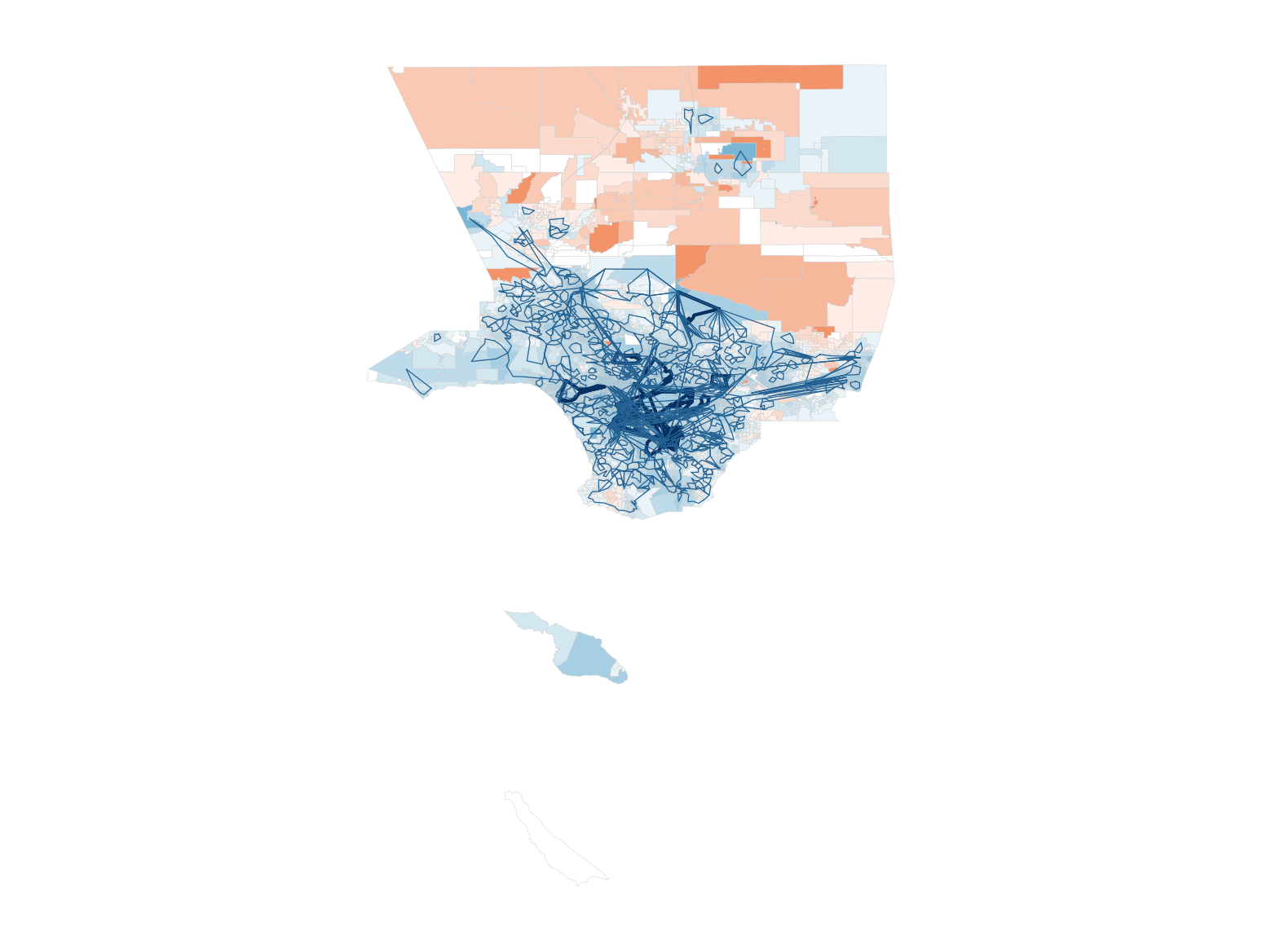}
\vspace{-.25 cm}
	\caption{\label{subfig:C_la_b}Adjacency complex}
\end{subfigure}
\begin{subfigure}[t]{\textwidth}
	\centering
	\includegraphics[width=.45\textwidth]{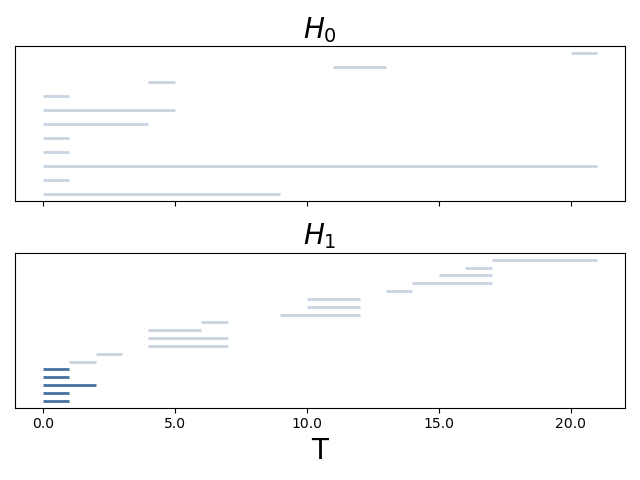}\quad\quad\quad
	\includegraphics[width=.3\textwidth, trim={2cm 0cm 2cm 1cm}, clip]{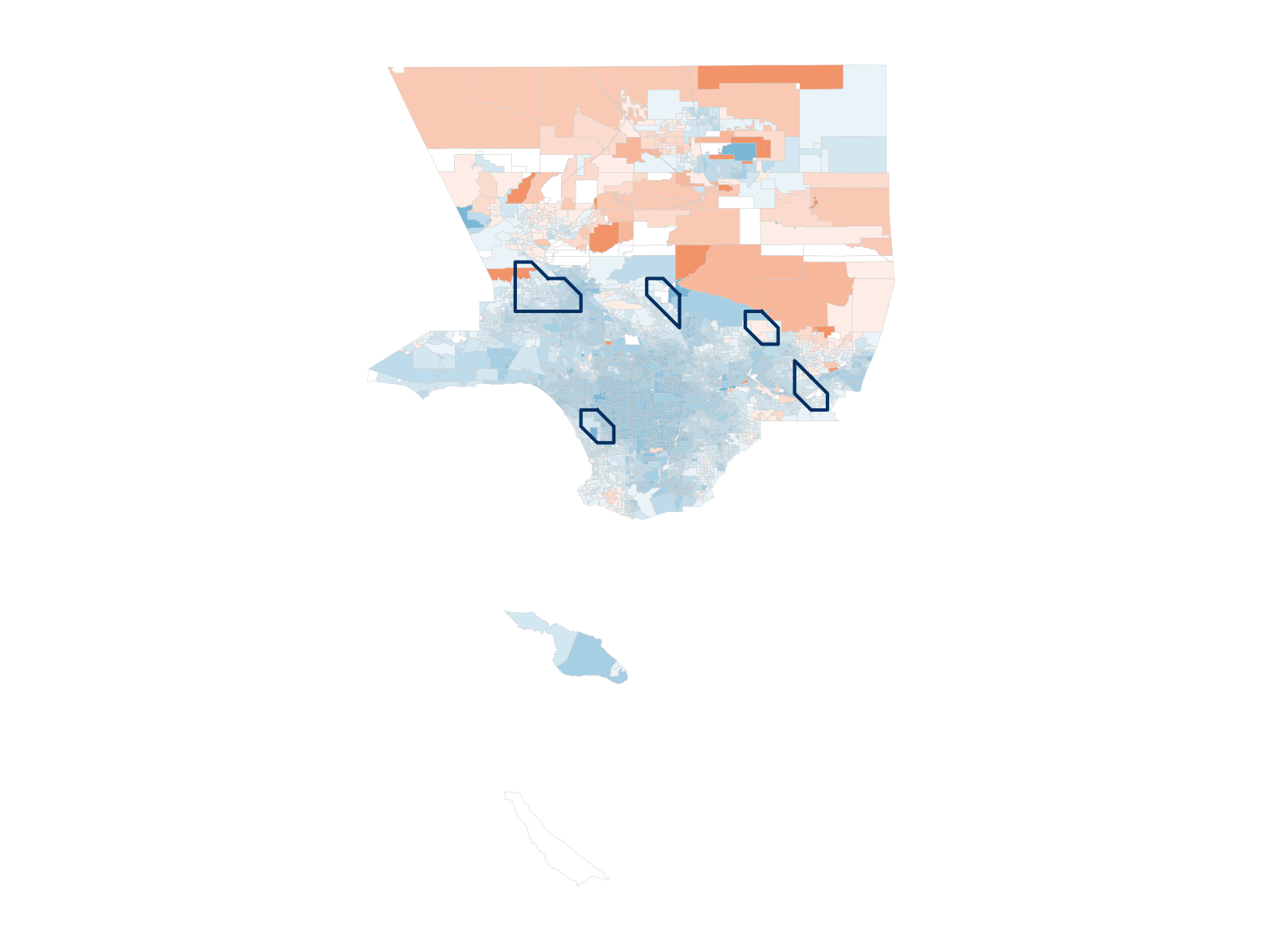} 
\vspace{-.5 cm}
	\caption{\label{subfig:C_la_c}Level-set complex}
\end{subfigure}
\vspace{-.5 cm}
\caption{Barcodes and generated loops for blue precincts in Los Angeles County. We again observe many non-long persistence features in the $H_1$ barcode of the alpha complex. This arises from the fact that the southern part of the county has a much higher population density than the northern part. There is also a single long-persistence feature; it crosses a red corridor, but does not surround a hole. The large number of precincts in this county makes it difficult to interpret many of the cycles in the highly populated precincts by eye, but the alpha complex includes several cycles that traverse red swaths of the map that do not appear to be holes, whereas neither the adjacency complex nor the level-set complex exhibit this behavior. 
}
\label{fig:C_labar} 
\end{figure}

%%%%%

\section{Complete Tables}
\label{app:tab}
In Table~\ref{tab:D_complex}, we give the computation times for the constructions of all computed simplicial complexes.

\begin{table}[!htbp]
\centering
\caption{Computation times for the constructions of our simplicial complexes.
}
\label{tab:D_complex}
\resizebox{13cm}{!} {
\begin{tabular}{l l l l l l l l l}
\toprule
\multirow{2}{*}{County}  & \multicolumn{2}{c}{VR} & \multicolumn{2}{c}{Alpha} & \multicolumn{2}{c}{Adjacency} & \multicolumn{2}{c}{Level set} \\
\cmidrule(lr){2-3}
\cmidrule(lr){4-5}
\cmidrule(lr){6-7}
\cmidrule(lr){8-9}

& C & T & C & T & C & T & C & T \\
\midrule
Alameda & -- & 0.191 s & 0.742 s & -- & 1.62 s & 0.0019 s & 4.97 s & 4.76 s \\
Alpine & 0.00169 s & 0.015 s & -- & -- & 0.00174 s & 0.000727 s & 12.3 s & 15.5 s \\
Amador & 0.000706 s & 0.0323 s & -- & -- & 0.00281 s & 0.00591 s & 5.18 s & 5.24 s \\
Calaveras & 0.00117 s & 0.0172 s & -- & -- & 0.000872 s & 0.00248 s & 9.66 s & 7.41 s \\
Colusa & 0.00097 s & 0.00251 s & -- & -- & 0.00184 s & 0.00175 s & 4.96 s & 6.31 s \\
Contra Costa & -- & 0.593 s & 0.468 s & -- & 0.619 s & 0.0033 s & 4.81 s & 5.12 s \\
Del Norte & 0.0011 s & 0.0187 s & -- & -- & 0.00265 s & 0.0039 s & 13.1 s & 10.6 s \\
El Dorado & 0.302 s & 182 s & -- & -- & 0.00363 s & 0.0905 s & 5.46 s & 5.62 s \\
Fresno & -- & -- & 0.143 s & 0.0952 s & 0.123 s & 0.102 s & 7.73 s & 8.54 s \\
Glenn & 0.00116 s & 0.0433 s & -- & -- & 0.000836 s & 0.00421 s & 5.45 s & 5.3 s \\
Humboldt & 43.7 s & 0.0214 s & -- & -- & 0.0309 s & 0.00644 s & 10.1 s & 10.6 s \\
Imperial & 20.7 s & 0.756 s & -- & -- & 0.0137 s & 0.00291 s & 9.29 s & 6.2 s \\
Inyo & 0.00102 s & 0.00329 s & -- & -- & 0.00112 s & 0.00215 s & 7.32 s & 8.02 s \\
Kern & -- & -- & 0.0737 s & 0.221 s & 0.109 s & 0.388 s & 3.34 s & 4.26 s \\
Kings & 0.93 s & 108 s & -- & -- & 0.104 s & 0.0847 s & 10.5 s & 17.4 s \\
Lake & 0.131 s & 0.0264 s & -- & -- & 0.00672 s & 0.00407 s & 10.7 s & 11.7 s \\
Lassen & 0.00195 s & 0.81 s & 0.000417 s & 0.0234 s & 0.00343 s & 0.0108 s & 10.9 s & 11.9 s \\
Los Angeles & -- & -- & 15.5 s & 0.133 s & 39.3 s & 0.0602 s & 9.96 s & 12.9 s \\
Madera & 0.0344 s & 0.13 s & -- & -- & 0.0046 s & 0.00399 s & 5.24 s & 6.03 s \\
Marin & -- & 0.00196 s & 0.0705 s & -- & 0.063 s & 0.000784 s & 8.46 s & 7.56 s \\
Mariposa & 0.0012 s & 0.0155 s & -- & -- & 0.00323 s & 0.00215 s & 5.32 s & 5.62 s \\
Mendocino & -- & 0.0317 s & 0.0857 s & -- & 0.0571 s & 0.00148 s & 10.3 s & 9.52 s \\
Merced & 489 s & 59.4 s & -- & -- & 0.0217 s & 0.0154 s & 6.68 s & 7.18 s \\
Modoc & $1.91 \times 10^{-6}$ s & 0.0112 s & -- & -- & $2.15 \times 10^{-6}$ s & 0.00271 s & 3.81 s & 4.37 s \\
Mono & 0.00116 s & 0.00194 s & -- & -- & 0.00152 s & 0.000919 s & 5.8 s & 5.78 s \\
Monterey & -- & 4.23 s & 0.272 s & -- & 0.0766 s & 0.00302 s & 5.49 s & 5.6 s \\
Napa & 655 s & 0.00569 s & -- & -- & 0.0478 s & 0.00115 s & 8.31 s & 8.47 s \\
Nevada & 0.168 s & 0.134 s & -- & -- & 0.00751 s & 0.00543 s & 3.24 s & 3.11 s \\
Orange & -- & -- & 0.844 s & 0.693 s & 1.1 s & 0.613 s & 8.1 s & 8.47 s \\
Placer & 0.736 s & -- & -- & 0.184 s & 0.0172 s & 0.553 s & 3.1 s & 3.35 s \\
Plumas & 0.00138 s & 0.0269 s & -- & -- & 0.00109 s & 0.00401 s & 4.65 s & 5.52 s \\
Riverside & -- & -- & 0.263 s & 0.422 s & 0.483 s & 0.554 s & 2.21 s & 1.99 s \\
Sacramento & -- & -- & 0.516 s & 0.0841 s & 12.3 s & 0.606 s & 8.48 s & 9.5 s \\
San Benito & 0.1 s & 0.00899 s & -- & -- & 0.00662 s & 0.00339 s & 6.14 s & 6.79 s \\
San Bernardino & -- & -- & 1.77 s & 0.833 s & 0.691 s & 0.476 s & 4.39 s & 5.25 s \\
San Diego & -- & -- & 1.63 s & 0.492 s & 3.13 s & 0.416 s & 6.11 s & 6.87 s \\
San Francisco & -- & 0 s & 0.353 s & -- & 0.707 s & 1.19e-06 s & 5.99 s & 5.13 s \\
San Joaquin & -- & -- & 0.0857 s & 0.0487 s & 0.108 s & 0.052 s & 8.22 s & 13.2 s \\
San Luis Obispo & 14.4 s & 4.45 s & -- & -- & 0.016 s & 0.0115 s & 3.61 s & 3.88 s \\
San Mateo & -- & 0.00359 s & 0.266 s & -- & 0.45 s & 0.00442 s & 8.28 s & 5.84 s \\
Santa Barbara & -- & 3.35 s & 0.0551 s & -- & 0.0612 s & 0.0101 s & 6.35 s & 7.36 s \\
Santa Cruz & -- & 0.0017 s & 0.0625 s & -- & 0.207 s & 0.00239 s & 5.7 s & 10.4 s \\
Shasta & 0.00118 s & 120 s & -- & -- & 0.00526 s & 0.0426 s & 3.9 s & 4.47 s \\
Sierra & 0.00342 s & 0.00752 s & -- & -- & 0.0046 s & 0.00311 s & 2.86 s & 3.3 s \\
Solano & 401 s & 3.59 s & -- & -- & 0.0438 s & 0.013 s & 6.24 s & 5.86 s \\
Sonoma & -- & 0.0299 s & 0.25 s & -- & 0.516 s & 0.0175 s & 6.26 s & 5.46 s \\
Stanislaus & 46.1 s & 52.1 s & -- & -- & 0.0173 s & 0.0325 s & 8.25 s & 9.41 s \\
Sutter & 0.00187 s & 0.244 s & -- & -- & 0.00494 s t& 0.0193 s & 8.75 s & 9.79 s \\
Tehama & $1.91 \times 10^{-6}$ s & 0.372 s & -- & -- & $3.1 \times 10^{-6}$ s & 0.0601 s & 2.32 s & 2.78 s \\
Trinity & 0.000981 s & 0.0038 s & -- & -- & 0.00276 s & 0.0225 s & 8.44 s & 8.86 s \\
Tulare & 3.57 s & -- & -- & 0.0515 s & 0.0562 s & 0.129 s & 4.81 s & 5.18 s \\
Tuolomne & 0.000928 s & 2.15 s & -- & -- & 0.00328 s & 0.0117 s & 4.76 s & 4.5 s \\
Yolo & 51.4 s & 0.0142 s & -- & -- & 0.0635 s & 0.00449 s & 5.92 s & 6.1 s \\
Yuba & 0.00109 s & 0.266 s & -- & -- & 0.00988 s & 0.0096 s & 7.97 s & 8.72 s \\
\end{tabular}
}
\end{table}

%%%%%

\section*{Acknowledgements}

We thank Moon Duchin, Joshua Gensler, Mike Hill, Stan Osher, Nina Otter, Bao Wang, and two anonymous referees for helpful comments on this project. We also thank Emilia Alvarez, Eion Blanchard, Austin Eide, Patrick Girardet, Everett Meike, Dmitriy Morozov, Justin Solomon, Courtney Thatcher, Jim Thatcher, and Maia Woluchem for insightful discussions.

%%%%%%

%\bibliographystyle{siamplain}
%\bibliography{ref5}

\begin{thebibliography}{10}

\bibitem{adamaszek_adams_2017}
{\sc M.~Adamaszek and H.~Adams}, {\em The vietoris--rips complexes of a
  circle}, Pacific Journal of Mathematics, 290 (2017), p.~1?40,
  \url{https://doi.org/10.2140/pjm.2017.290.1}.

\bibitem{Adams:2017:PIS:3122009.3122017}
{\sc H.~Adams, T.~Emerson, M.~Kirby, R.~Neville, C.~Peterson, P.~Shipman,
  S.~Chepushtanova, E.~Hanson, F.~Motta, and L.~Ziegelmeier}, {\em Persistence
  images: A stable vector representation of persistent homology}, Journal of
  Machine Learning Research, 18 (2017), pp.~218--252.

\bibitem{Bajardi2015}
{\sc P.~Bajardi, M.~Delfino, A.~Panisson, G.~Petri, and M.~Tizzoni}, {\em
  Unveiling patterns of international communities in a global city using mobile
  phone data}, EPJ Data Science, 4 (2015), p.~3.

\bibitem{banman2018}
{\sc A.~Banman and L.~Ziegelmeier}, {\em Mind the gap: {A} study in global
  development through persistent homology}, in Research in Computational
  Topology, E.~W. Chambers, B.~T. Fasy, and L.~Ziegelmeier, eds., Springer
  International Publishing, Cham, Switzerland, 2018, pp.~125--144.

\bibitem{DBLP:journals/corr/abs-1803-02857}
{\sc R.~Barnes and J.~Solomon}, {\em Gerrymandering and compactness:
  {I}mplementation flexibility and abuse}, arXiv:1803.02857,  (2018).

\bibitem{10.1007/978-3-662-44199-2_24}
{\sc U.~Bauer, M.~Kerber, J.~Reininghaus, and H.~Wagner}, {\em {\sc Phat} ---
  {P}ersistent homology algorithms toolbox}, in Mathematical Software --- ICMS
  2014, H.~Hong and C.~Yap, eds., Berlin, Heidelberg, 2014, Springer Berlin
  Heidelberg, pp.~137--143.

\bibitem{bendich2016}
{\sc P.~Bendich, J.~S. Marron, E.~Miller, A.~Pieloch, and S.~Skwerer}, {\em
  Persistent homology analysis of brain artery trees}, The Annals of Applied
  Statistics, 10 (2016), pp.~198--218.

\bibitem{bobrowski2017}
{\sc O.~Bobrowski, S.~Mukherjee, and J.~E. Taylor}, {\em Topological
  consistency via kernel estimation}, Bernoulli, 23 (2017), pp.~288--328.

\bibitem{Bubenik:2015:STD:2789272.2789275}
{\sc P.~Bubenik}, {\em Statistical topological data analysis using persistence
  landscapes}, Journal of Machine Learning Research, 16 (2015), pp.~77--102.

\bibitem{Carlsson2008}
{\sc G.~Carlsson, T.~Ishkhanov, V.~de~Silva, and A.~Zomorodian}, {\em On the
  local behavior of spaces of natural images}, International Journal of
  Computer Vision, 76 (2008), pp.~1--12.

\bibitem{curto2016}
{\sc C.~Curto}, {\em What can topology tell us about the neural code?},
  Bulletin of the American Mathematical Society, 54 (2017), pp.~63--78.

\bibitem{10.7554/eLife.03476}
{\sc Y.~Dabaghian, V.~L. Brandt, and L.~M. Frank}, {\em Reconceiving the
  hippocampal map as a topological template}, eLife, 3 (2014), p.~e03476.

\bibitem{2018arXiv180805860D}
{\sc M.~{Duchin} and B.~E. {Tenner}}, {\em {Discrete geometry for electoral
  geography}}, arXiv:1808.05860,  (2018).

\bibitem{edelsbrunner2010}
{\sc H.~Edelsbrunner and J.~Harer}, {\em Computational Topology: An
  Introduction}, American Mathematical Society, Providence, RI, USA, 2010.

\bibitem{1056714}
{\sc H.~Edelsbrunner, D.~Kirkpatrick, and R.~Seidel}, {\em On the shape of a
  set of points in the plane}, IEEE Transactions on Information Theory, 29
  (1983), pp.~551--559.

\bibitem{Emmett:2016:MTC:2954721.2954838}
{\sc K.~Emmett, B.~Schweinhart, and R.~Rabadan}, {\em Multiscale topology of
  chromatin folding}, in Proceedings of the 9th EAI International Conference on
  Bio-inspired Information and Communications Technologies (Formerly
  BIONETICS), BICT'15, ICST, Brussels, Belgium, Belgium, 2016, ICST (Institute
  for Computer Sciences, Social-Informatics and Telecommunications
  Engineering), pp.~177--180.

\bibitem{Gameiro2015}
{\sc M.~Gameiro, Y.~Hiraoka, S.~Izumi, M.~Kramar, K.~Mischaikow, and V.~Nanda},
  {\em A topological measurement of protein compressibility}, Japan Journal of
  Industrial and Applied Mathematics, 32 (2015), pp.~1--17.

\bibitem{ghrist2008}
{\sc R.~Ghrist}, {\em {Barcodes: The persistent topology of data}}, Bulletin of
  the American Mathematical Society, 45 (2008), pp.~61--75.

\bibitem{GIBOU201882}
{\sc F.~Gibou, R.~Fedkiw, and S.~Osher}, {\em A review of level-set methods and
  some recent applications}, Journal of Computational Physics, 353 (2018),
  pp.~82 -- 109,
  \url{https://doi.org/https://doi.org/10.1016/j.jcp.2017.10.006},
  \url{http://www.sciencedirect.com/science/article/pii/S0021999117307441}.

\bibitem{Giusti2016}
{\sc C.~Giusti, R.~Ghrist, and D.~S. Bassett}, {\em Two's company, three (or
  more) is a simplex}, Journal of Computational Neuroscience, 41 (2016),
  pp.~1--14.

\bibitem{hatcher2002algebraic}
{\sc A.~Hatcher}, {\em Algebraic Topology}, Cambridge University Press,
  Cambridge, UK, 2002.

\bibitem{qgis}
{\sc D.~P. Humphreys, M.~R. McGuirl, M.~Miyagi, and A.~J. Blumberg}, {\em {QGIS
  2.18.17}: {A} free and open source geographic information system}, 2016,
  \url{https://qgis.org/en/site/}.

\bibitem{blumberg2018}
{\sc D.~P. Humphreys, M.~R. McGuirl, M.~Miyagi, and A.~J. Blumberg}, {\em Fast
  estimation of recombination rates using topological data analysis}, Genetics,
  211 (2019), pp.~1191--1204.

\bibitem{Ignacio2019}
{\sc P.~S.~P. Ignacio and I.~K. Darcy}, {\em Tracing patterns and shapes in
  remittance and migration networks via persistent homology}, EPJ Data Science,
  8 (2019), p.~1.

\bibitem{Kanari2016QuantifyingTI}
{\sc L.~Kanari, P.~D\l{}otko, M.~Scolamiero, R.~Levi, J.~C. Shillcock, K.~Hess,
  and H.~Markram}, {\em Quantifying topological invariants of neuronal
  morphologies}, arXiv:1603.08432,  (2016).

\bibitem{kerber2013}
{\sc M.~Kerber and R.~Sharathkumar}, {\em Approximate \v{C}ech complex in low
  and high dimensions}, arXiv:1307.3272,  (2013).

\bibitem{kovacev2016}
{\sc V.~Kovacev-Nikolic, P.~Bubenik, D.~Nikolic, and G.~Heo}, {\em Using
  persistent homology and dynamical distances to analyze protein binding},
  Statistical Applications in Genetics and Molecular Biology, 15 (2016),
  pp.~19--38.

\bibitem{NIPS2015_5887}
{\sc R.~Kwitt, S.~Huber, M.~Niethammer, W.~Lin, and U.~Bauer}, {\em Statistical
  topological data analysis --- {A} kernel perspective}, in Advances in Neural
  Information Processing Systems 28, C.~Cortes, N.~D. Lawrence, D.~D. Lee,
  M.~Sugiyama, and R.~Garnett, eds., Curran Associates, Inc., 2015,
  pp.~3070--3078.

\bibitem{10.1371/journal.pone.0192120}
{\sc D.~Lo and B.~Park}, {\em Modeling the spread of the {Z}ika virus using
  topological data analysis}, {PLoS ONE}, 13 (2018), p.~e0192120.

\bibitem{10.3389/fnsys.2016.00085}
{\sc L.-D. Lord, P.~Expert, H.~M. Fernandes, G.~Petri, T.~J. Van~Hartevelt,
  F.~Vaccarino, G.~Deco, F.~Turkheimer, and M.~L. Kringelbach}, {\em Insights
  into brain architectures from the homological scaffolds of functional
  connectivity networks}, Frontiers in Systems Neuroscience, 10 (2016), p.~85.

\bibitem{gudhi:FilteredComplexes}
{\sc C.~Maria}, {\em Filtered complexes}, in {\sc Gudhi} User and Reference
  Manual, {\sc Gudhi} Editorial Board, 2015,
  \url{http://gudhi.gforge.inria.fr/doc/latest/group__simplex__tree.html}.

\bibitem{gudhi:RipsComplex}
{\sc C.~Maria, P.~D\l{}otko, V.~Rouvreau, and M.~Glisse}, {\em Rips complex},
  in {\sc Gudhi} User and Reference Manual, {\sc Gudhi} Editorial Board, 2016,
  \url{http://gudhi.gforge.inria.fr/doc/latest/group__rips__complex.html}.

\bibitem{newman2018}
{\sc M.~E.~J. Newman}, {\em Networks}, Oxford University Press, Oxford, UK,
  2~ed., 2018.

\bibitem{osher2003}
{\sc S.~Osher and R.~Fedkiw}, {\em Level Set Methods and Dynamic Implicit
  Surfaces}, vol.~153 of AMS, Springer-Verlag, Berlin, Germany, 2003.

\bibitem{OSHER198812}
{\sc S.~Osher and J.~A. Sethian}, {\em Fronts propagating with
  curvature-dependent speed: {A}lgorithms based on {H}amilton--{J}acobi
  formulations}, Journal of Computational Physics, 79 (1988), pp.~12--49.

\bibitem{otter2017}
{\sc N.~Otter, M.~A. Porter, U.~Tillmann, P.~Grindrod, and H.~A. Harrington},
  {\em {A roadmap for the computation of persistent homology}}, European
  Physical Journal --- Data Science, 6 (2017), p.~17.

\bibitem{lia2018}
{\sc L.~Papadopoulos, M.~A. Porter, K.~E. Daniels, and D.~S. Bassett}, {\em
  Network analysis of particles and grains}, Journal of Complex Networks, 6
  (2018), pp.~485--565.

\bibitem{dotko2016}
{\sc M.~W. Reimann, M.~Nolte, M.~Scolamiero, K.~Turner, R.~Perin, G.~Chindemi,
  P.~D\l{}otko, R.~Levi, K.~Hess, and H.~Markram}, {\em Cliques of neurons
  bound into cavities provide a missing link between structure and function},
  Frontiers in Computational Neuroscience, 11 (2017), p.~48.

\bibitem{Reininghaus2015ASM}
{\sc J.~Reininghaus, S.~Huber, U.~Bauer, and R.~Kwitt}, {\em A stable
  multi-scale kernel for topological machine learning}, in 2015 IEEE Conference
  on Computer Vision and Pattern Recognition (CVPR), 06 2015, pp.~4741--4748.

\bibitem{flow2019}
{\sc J.~W. Rocks, A.~J. Liu, and E.~Katifori}, {\em The topological basis of
  function in flow networks}, arXiv:1901.00822,  (2019).

\bibitem{gudhi:AlphaComplex}
{\sc V.~Rouvreau}, {\em Alpha complex}, in {\sc Gudhi} User and Reference
  Manual, {\sc Gudhi} Editorial Board, 2015,
  \url{http://gudhi.gforge.inria.fr/doc/latest/group__alpha__complex.html}.

\bibitem{gudhi:cython}
{\sc V.~Rouvreau}, {\em Cython interface}, in {\sc Gudhi} User and Reference
  Manual, {\sc Gudhi} Editorial Board, 2016,
  \url{http://gudhi.gforge.inria.fr/python/latest/}.

\bibitem{schleuss2016}
{\sc J.~Schleuss, J.~Fox, and P.~Krishnakumar}, {\em California 2016 election
  precinct maps}.
\newblock
  \url{https://github.com/datadesk/california-2016-election-precinct-maps},
  2016.

\bibitem{speidel2018}
{\sc L.~Speidel, H.~A. Harrington, S.~J. Chapman, and M.~A. Porter}, {\em
  Topological data analysis of continuum percolation with disks}, Physical
  Review E, 98 (2018), p.~012318.

\bibitem{stolz-brexit}
{\sc B.~J. Stolz, H.~A. Harrington, and M.~A. Porter}, {\em The topological
  ``shape'' of {B}rexit}, arXiv:1610.00752,  (2016).

\bibitem{doi:10.1063/1.4978997}
{\sc B.~J. Stolz, H.~A. Harrington, and M.~A. Porter}, {\em Persistent homology
  of time-dependent functional networks constructed from coupled time series},
  Chaos, 27 (2017), p.~047410.

\bibitem{taylor2015}
{\sc D.~Taylor, F.~Klimm, H.~A. Harrington, M.~Kram{\'{a}}r, K.~Mischaikow,
  M.~A. Porter, and P.~J. Mucha}, {\em {Topological data analysis of contagion
  maps for examining spreading processes on networks}}, Nature Communications,
  6 (2015), p.~7723.

\bibitem{gudhi:urm}
{\sc {The {\sc Gudhi} Project}}, {\em {\sc Gudhi} User and Reference Manual},
  {\sc Gudhi} Editorial Board, 2015,
  \url{http://gudhi.gforge.inria.fr/doc/latest/}.

\bibitem{Vietoris1927}
{\sc L.~Vietoris}, {\em {\"U}ber den h{\"o}heren zusammenhang kompakter
  r{\"a}ume und eine klasse von zusammenhangstreuen abbildungen}, Mathematische
  Annalen, 97 (1927), pp.~454--472.

\bibitem{doi:10.1002/cnm.2655}
{\sc K.~Xia and G.-W. Wei}, {\em Persistent homology analysis of protein
  structure, flexibility, and folding}, International Journal for Numerical
  Methods in Biomedical Engineering, 30 (2014), pp.~814--844.

\bibitem{YOO20161}
{\sc J.~Yoo, E.~Y. Kim, Y.~M. Ahn, and J.~C. Ye}, {\em Topological persistence
  vineyard for dynamic functional brain connectivity during resting and gaming
  stages}, Journal of Neuroscience Methods, 267 (2016), pp.~1--13.

\bibitem{doi:10.1093/bib/bbs077}
{\sc W.~Zhou and H.~Yan}, {\em Alpha shape and {D}elaunay triangulation in
  studies of protein-related interactions}, Briefings in Bioinformatics, 15
  (2014), pp.~54--64.

\bibitem{Zhu:2016:SMP:3060832.3060964}
{\sc X.~Zhu, A.~Vartanian, M.~Bansal, D.~Nguyen, and L.~Brandl}, {\em
  Stochastic multiresolution persistent homology kernel}, in Proceedings of the
  Twenty-Fifth International Joint Conference on Artificial Intelligence,
  IJCAI'16, AAAI Press, 2016, pp.~2449--2455.

\bibitem{Zomorodian_2010}
{\sc A.~Zomorodian}, {\em Fast construction of the {V}ietoris--{R}ips complex},
  Computers {\&} Graphics, 34 (2010), pp.~263--271.

\bibitem{Zomorodian2005}
{\sc A.~Zomorodian and G.~Carlsson}, {\em Computing persistent homology},
  Discrete {\&} Computational Geometry, 33 (2005), pp.~249--274.

\end{thebibliography}

%%%%%

\end{document}